\documentclass[prb,twocolumn,aps,groupedaddress,nofootinbib]{revtex4}

\usepackage{graphicx,amsfonts,amssymb,amsmath,hyperref,hypcap,enumerate}
\usepackage{slashed}
\usepackage{amsmath,amssymb,bm,graphicx,dsfont}
\usepackage[latin9]{inputenc}
\usepackage{amsmath}
\usepackage{slashed}
\usepackage{dsfont}
\usepackage{amstext}
\usepackage{amssymb}
\usepackage{amsbsy}
\usepackage{amsthm}
\usepackage{epsfig}
\usepackage{graphicx}
\usepackage{bbm}
\usepackage{hyperref}
\usepackage{color}
\usepackage{slashed}
\newcommand{\up}{\uparrow}
\newcommand{\down}{\downarrow}

\renewcommand{\(}{\left(}
\renewcommand{\)}{\right)}

\newcommand{\T}{\mathcal{T}}
\newcommand{\C}{\mathcal{C}}
\newcommand{\CT}{\mathcal{CT}}

\newcommand{\mcp}{\mathcal{P}}
\renewcommand{\P}{\mathcal{P}}
\newcommand{\be}{\begin{equation}}
\newcommand{\ba}{\begin{aligned}}
\newcommand{\ea}{\end{aligned}}
\newcommand{\ee}{\end{equation}}
\newcommand{\bea}{\begin{eqnarray}}
\newcommand{\eea}{\end{eqnarray}}
\newcommand{\beq}{\begin{equation}}
\newcommand{\eeq}{\end{equation}}
\newcommand{\beqn}{\begin{eqnarray}}
\newcommand{\eeqn}{\end{eqnarray}}

\newcommand{\p}{\partial}

\newcommand{\Tr}{{\rm \, Tr\,}}
\newcommand{\Z}{\mathbb{Z}}

\newcommand{\upa}{\uparrow}
\newcommand{\doa}{\downarrow}

\newcommand{\tp}{\otimes}
\newcommand{\unit}{\mathbf{1}}
\newcommand{\mt}{\mapsto}
\newcommand{\bfx}{\bm{x}}

\newcommand{\psp}{{PSp(N_f)}}

\definecolor{rose}{rgb}{1.0, 0.01, 0.24}
\definecolor{blueish}{rgb}{0.2, 0.5, 0.85}

\newcommand{\ch}{{\rm ch}}

\begin{document}
\title{Landau ordering phase transitions beyond the Landau paradigm}
\author{Zhen Bi}
\thanks{These two authors contributed equally.}
\author{Ethan Lake}
\thanks{These two authors contributed equally.}
\author{T. Senthil}
\affiliation{Department of Physics, Massachusetts Institute of Technology, Cambridge, MA, 02139}
\begin{abstract}
Continuous phase transitions associated with the onset of a spontaneously broken symmetry are thought to be successfully described by the Landau-Ginzburg-Wilson-Fisher theory of fluctuating order parameters. In this work we show that such transitions can admit  new universality classes which cannot be understood in terms of a theory of order parameter fluctuations.  We  explicitly demonstrate  continuous time reversal symmetry breaking quantum phase transitions of  $3+1$-D bosonic systems described by  critical theories expressed in terms of a deconfined gauge theory with massless Dirac fermions instead of the fluctuating Ising order parameter. We dub such phase transitions ``Landau transitions beyond Landau description" (LBL).  A key feature of our examples is that the stability of the LBL fixed points requires a crucial global symmetry, which is non-anomalous, unbroken, and renders no symmetry protected topological phase throughout the phase diagram. Despite this, there are elementary critical fluctuations of the phase transition that transform projectively  under this symmetry group. We also construct examples of other novel quantum critical phenomena, notably a continuous Landau-forbidden deconfined critical point between two Landau-allowed phases in $3+1$-D.

\end{abstract}
\maketitle

\section{Introduction} 
The standard example of continuous equilibrium  phase transitions is that associated with the onset of  a broken symmetry. The symmetry breaking is captured by a Landau order parameter. 
The corresponding critical phenomena   are then  associated with the long wavelength long time fluctuations of this order parameter. These fluctuations can be described by a continuum quantum field theory written in terms of this order parameter field.  This paradigm (developed primarily by Landau, Ginzburg, Wilson, and Fisher (LGWF)) in combination with Renormalization Group (RG) methods provides a powerful and remarkably successful framework for describing phase transitions, both classical and quantum\cite{sachdevbook}.  

The LGWF paradigm is known to fail in a few different situations. First,  one or both phases may have order not captured by a Landau order parameter (eg, quantum Hall or other topological phases). In this case it is of course natural that the critical point is not described by an order parameter-based theory.  Second, more surprisingly, it is known that there are Landau-forbidden second order quantum phase transitions between two phases that themselves are Landau allowed\cite{2dDQCP1,2dDQCP2,2dDQCPjps,2dDQCPprb,kaul2013bridging,sandvik2010}.
Such phase transitions are more naturally described in terms of fractionalized degrees of freedom which rear their heads only at the critical point but are absent (confined) in either of the two phases. Hence they are dubbed deconfined quantum critical points.  

\begin{figure}
\begin{center}
\includegraphics[width=0.9\linewidth]{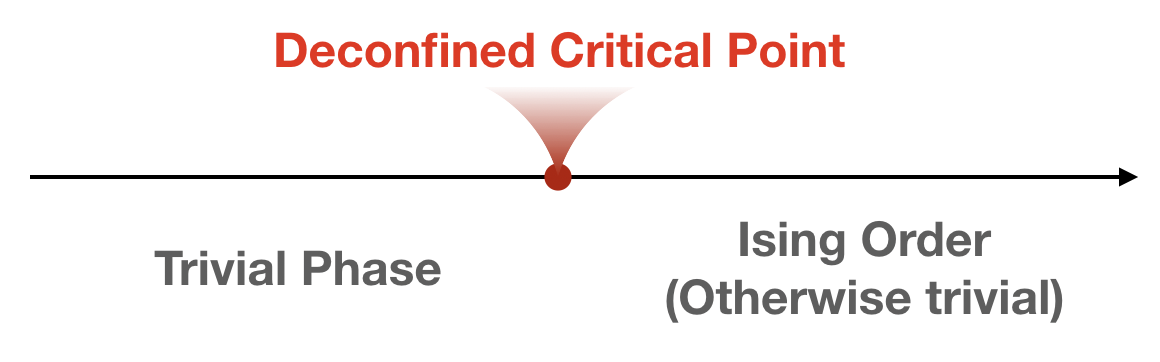}
\caption{A schematic plot of the phase diagram. In addition to the standard Ising universality class the transition can also occur through a distinct `deconfined critical' universality class.}
\label{fig:PD0}
\end{center}
\end{figure}

\begin{figure}
\begin{center}
\includegraphics[width=0.9\linewidth]{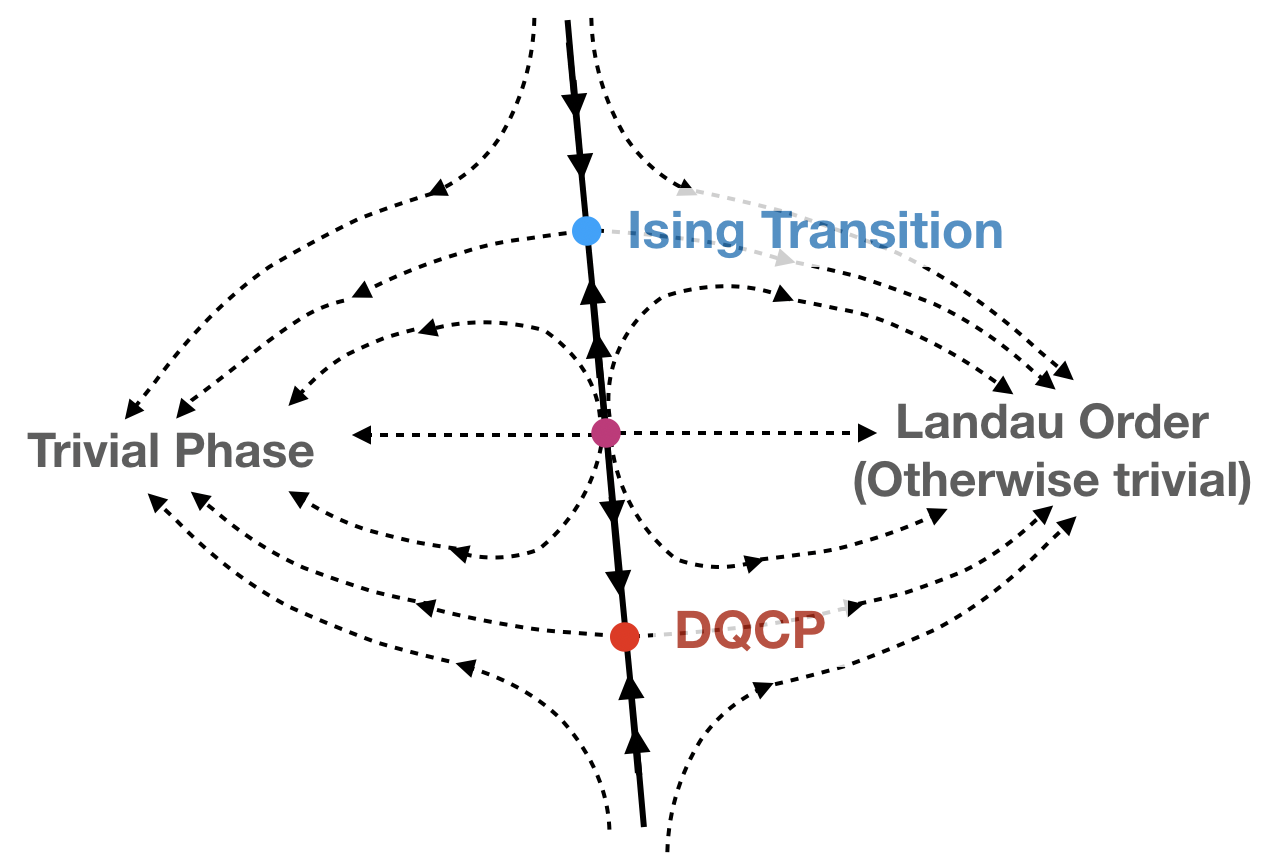}
\caption{A schematic plot of the RG flow showing both the Ising and the alternate deconfined critical fixed points for the same phase transition.}
\label{fig:sPD}
\end{center}
\end{figure}

In this paper we demonstrate the breakdown of the LGWF framework even for a standard quantum phase transition between a trivial phase and a broken symmetry phase (which is otherwise also trivial).  The corresponding transition is allowed to be in the standard universality class described by LGWF theory based on the order parameter field. However, we show the existence also of a different deconfined quantum critical fixed point with emergent fractionalized excitations. A schematic phase diagram is shown in Fig. \ref{fig:PD0}. Examples of multiple universality classes for the same phase transition were discussed extensively in previous work\cite{BS} by two of the authors. While the previous examples focused on topological phase transitions, here our focus is on Landau ordering transitions (see Fig. \ref{fig:sPD} for an illustration of the associated renormalization group flow diagram showing both critical fixed points).  At the new fixed points we find the physics cannot be described in terms of order parameter fluctuations alone. Thus we have a situation where a Landau allowed phase transition is not necessarily described within the Landau paradigm. We dub such a phase transition ``Landau Beyond Landau" (LBL). 

The possibility that the LGWF framework might fail at some Landau ordering transitions has long been suggested in the context of experiments on heavy electron metals\cite{sietal,tsmvss04,tsssmv05, HFRMP,Coleman}. However, the complications of dealing with issues related to the metallic Fermi surface has stymied progress on many basic questions. The concrete examples we construct are in a much simpler setting, and establish the point of principle that order parameter fluctuations might distract from the true criticality even for phase transitions between trivial and Landau ordered phases. 

Our basic example  is a  $3+1$-D system of bosons with a global symmetry group $G \times \mathcal{T}$, 
where $G$ is a continuous unitary group and $\mathcal{T}$ is time reversal (and hence anti-unitary).  The phase transition occurs between a phase with unbroken $G \times \mathcal{T}$ symmetry to a phase which spontaneously breaks $\mathcal{T}$ but preserves $G$.  All excitations in either phase are gapped.  Further, neither phase has any exotic order (no intrinsic topological order or  Symmetry Protected Topological order).  Then the onset of spontaneous $\mathcal{T}$ symmetry breaking could happen through the conventional $3+1$-D Ising universality class (which is Gaussian with a marginally irrelevant perturbation).  Here the $G$ symmetry does not act on the low energy critical degrees of freedom. However we will also find an alternate route through a `deconfined critical' universality class with emergent fermions and gauge fields.  At this deconfined critical fixed point, the symmetry $G$ acts non-trivially on the critical degrees of freedom. Consequently local ({\em i.e} gauge invariant) operators that transform under $G$ have power law correlations unlike at the conventional Ising universality class. 

Though the ``Landau beyond Landau" transitions are the main focus of this paper, we will also briefly describe some other possible novel quantum phase transitions. Particularly interesting are concrete examples (see Apendix \ref{app:3dsu2b}) of continuous Landau-forbidden phase transitions between Landau allowed phases in $3+1$-D. We describe deconfined quantum critical points for such transitions. These examples generalize similar phenomena known in $2+1$-D, and settle the matter-of-principle question on whether such continuous quantum phase transitions can occur in $3+1$-D.

\section{An Ising transition beyond Landau description}
To construct our examples we follow the same strategy as in Ref. \onlinecite{BS}, and interpret some known conformal field theories as quantum critical points.
\subsection{Parton field theory and phase diagram}
We begin by looking at  $SU(2)$ gauge theory in $3+1$-D with $N_f$ Dirac fermions in the fundamental representation. The Lagrangian is
\beq
\mathcal{L}=\sum_{j=1}^{N_f}i\bar{\psi}_j\gamma^\mu(\partial_\mu-ia_\mu)\psi_j -m\bar{\psi}\psi+\frac{1}{4g^2}\Tr f_{\mu\nu}^2+...,
\label{eq:L}
\eeq
where $a_\mu$ and $f_{\mu\nu}$ are the dynamical $SU(2)$ gauge field and field strength respectively, and $\psi$'s are Dirac fermions that are in the spin-$\frac{1}{2}$ representation of the $SU(2)$ gauge group. As stressed in previous papers, despite appearances this theory is intrinsically bosonic: all gauge invariant operators (\textit{e.g.} baryons) are bosons. We will regard this Lagrangian as an intermediate energy scale description of a microscopic system of these gauge invariant bosons (possibly on a lattice).  Stated differently, the gauge theory can be viewed as a parton theory of the underlying gauge invariant bosonic degrees of freedom. 

Global symmetries of the theory will be important for our discussion. Consider the symmetry of the parton theory at generic mass $m$. First, there is a continuous global symmetry (see, eg, Ref. \onlinecite{BS}, and App. \ref{app:sym}) $G=\psp=Sp(N_f)/\Z_2$ corresponding to flavor rotations of the fermions modulo gauge transformations. In addition, the theory also preserves the discrete symmetries of time reversal $\T$ and parity $\mathcal{P}$.\footnote{There is actually no independent notion of charge conjugation $\C$ in these theories; the action of what we might naively refer to as $\C$ is already incorporated into the $PSp(N_f)$ symmetry.}  A detailed discussion of the global symmetries is given in App. \ref{app:sym}. The action of $\mathcal{T}$ and $\mathcal{P}$  on the fermions is
\bea
\mathcal{T}&:& \psi(\bm{x},t)\rightarrow \gamma_0\gamma_5\psi^\dagger(\bm{x},-t),\ \ \ i\rightarrow-i, 
\label{eq:T} \\
\mathcal{P}&:& \psi(\bm{x},t)\rightarrow \gamma_0\psi(-\bm{x},t) 
\eea
Notice the $PSp(N_f)$ together with the $\mathcal{T}$ and $\mathcal{P}$ symmetries are enough to prohibit all fermion bilinear terms other than $m\bar{\psi}\psi$.  Note also that we do not assume Lorentz symmetry, which is not appropriate as an exact symmetry in condensed matter systems. An explicit lattice model of low energy Dirac fermions with $\mathcal{T}$ and $\mathcal{P}$ symmetries is presented in App. \ref{app:lattice}. We will describe the role of the continuous symmetry group later. 

\begin{figure}
\begin{center}
\includegraphics[width=0.8\linewidth]{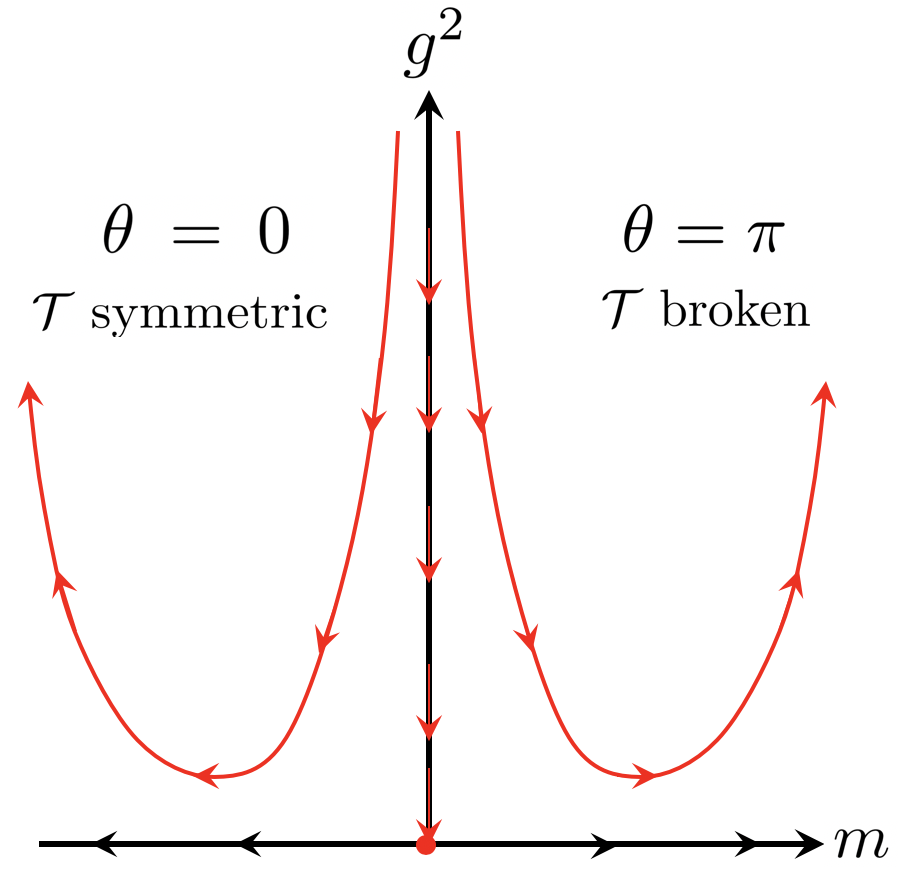}
\caption{ RG flow and phase diagram of the parton theory as a function of the fermion mass. For $m<0$, we are free to choose a regularization such that the gauge theory has trivial $\theta$ term and thus flows to a trivial confined phase. On the contrary, for $m>0$, at low energy the gauge theory has $\theta=\pi$, which leads to a  flow to a confined phase  that spontaneously breaks the $\mathcal{T}$ symmetry.}
\label{fig:PD}
\end{center}
\end{figure}

Ref. \onlinecite{BS} studied this theory for $N_f \in 2\mathbb{Z}$ to find examples of quantum critical points between trivial and Symmetry Protected Topological (SPT) phases of the bosons. Here we focus on the case of $N_f\in 2\mathbb{Z}+1$.  
We examine the phase diagram of the theory as a function of the fermion mass. At the massless point, with large enough flavors of fermions, the gauge theory is either in a strongly coupled\cite{BZnpb,BZprl,Shrock2013,Shrock2011} or non-interacting conformal field theory in the infrared (IR). We will interpret this as a critical point separating two gapped phases. For simplicity let us specialize to  $N_f$ large enough that  the gauge coupling is (marginally) irrelevant and the theory is IR free. When  a non-zero fermion mass is turned on, the RG flow of the gauge coupling (see Fig. \ref{fig:PD}) will turn around at a scale given by the mass. At lower energy scales we can  integrate out the fermions to obtain a pure gauge theory. 
At long distances, apart from the usual Yang-Mills term this induced effective action may have  a theta term
\beq
\mathcal{L}_{eff}=\frac{1}{4g^2}\Tr f_{\mu\nu}^2+\frac{\theta}{16\pi^2}\Tr \epsilon^{\mu\nu\lambda\rho}f_{\mu\nu}f_{\lambda\rho}+...
\eeq
The difference of the $\theta$-angle\footnote{This can be explicitly shown by calculating the ratio of the partition function between positive and negative mass with a fixed background gauge field and using the index theorem; see Ref. \onlinecite{BS} for details.} for $m>0$ and $m<0$ is $\Delta\theta=N_f\pi$. We are free to choose a regularization such that the $m<0$ side has trivial $\theta$-angle. In this phase, we have a pure $SU(2)$ gauge theory which enters a featureless confined phase in the infrared limit. This phase has a gap to excitations, preserves all the global symmetries, and has no topological order (even of the SPT kind). 
On the other side, $m>0$, the $SU(2)$ gauge theory has a non-trivial $\theta=N_f\pi$, which is equivalent to $\pi$ as $N_f$ is odd and $\theta$ is $2\pi$ periodic. 

The infrared dynamics of such an $SU(2)$ gauge theory is not fully understood. However, a variety of  reasons, which we review below, suggest a confining phase with spontaneously broken $\mathcal{T}$ and $\mathcal{P}$ symmetry as a promising candidate\cite{Witten, zohar2017, YMtheta,witten1980}.  In this section we will proceed by assuming this is the case, and show that we are then lead to the promised non-Landau description of a Landau ordering transition.  We will later refine our discussion in two ways. First we will generalize the theory to gauge groups $Sp(N_c)$ and $SU(N_c)$ where for large-$N_c$ it is known with more confidence that the pure gauge theory at $\theta = \pi$ enters a confined $\mathcal{T}$ and $\mathcal{P}$  broken phase. These generalizations will provide concrete examples of Landau ordering transitions beyond the Landau paradigm. A second refinement is to return to the $SU(2)$ gauge theory and allow for the possibility that the ground state at $\theta = \pi$ preserves $\mathcal{T}$ and $\mathcal{P}$ symmetries. Interesting  possible alternate ground states were proposed in Ref. \onlinecite{zohar2017}: the two we will focus on are deconfined $U(1)$ or $\Z_2$ gauge theories. From a condensed matter perspective these correspond to $U(1)$ or $\Z_2$ quantum spin liquid phases of the underlying boson system. We will show that in these scenarios the theory in Eq. \ref{eq:L} describes novel quantum critical points between such quantum spin liquids and a trivial gapped phase.

Assuming therefore for now that pure $SU(2)$ gauge theory at $\theta = \pi$ enters a confined phase with broken $\mathcal{T}$ symmetry, we conclude that this is also the fate for the $m > 0$ side of the theory in Eq. \ref{eq:L}. Thus the $m = 0$ IR-free theory will describe a critical point between a symmetry preserving trivial phase (for $m <0$) and a $\mathcal{T}$ breaking phase (for $m > 0$).  Clearly both phases preserve the global $PSp(N_f)$ symmetry.  Furthermore, neither phase has any intrinsic topological order or any gapless excitations.  This transition is apparently allowed within the usual LGWF framework and should be in the $3+1$-D Ising universality class; instead our critical theory is a deconfined gauge theory which is clearly distinct from the Ising universality class. 

However, there is a potential loophole to be closed before we can reach this conclusion. We still need to examine if either phase has SPT order associated with any of the unbroken global symmetries. If such order were present, the transition could clearly be in a distinct universality class from that in the LGWF framework. We will turn to this issue momentarily.


\subsection{The role of the global symmetry} 
First consider the continuous global symmetry $G= Sp(N_f)/\Z_2 =\psp$ (for an explanation of how this arises, see App. \ref{app:sym}).  This symmetry is not anomalous and is unbroken at any point of the phase diagram. However, this symmetry is important for our story.   If  we allow this symmetry to be explicitly broken, then we can add additional fermion bilinear terms (for instance different masses for different fermion flavors) to the theory, which are strongly relevant at the critical point. These relevant perturbations either lead the system to a gapped intermediate phase or lead the renormalization group flow to a new fixed point -- possibly the Ising fixed point if the $\psp$ is completely broken.  Thus the contemplated  nontrivial phase transition is protected by this additional global symmetry. Similar phenomena are also observed in recent developments on $1+1$-D deconfined quantum critical points\cite{1dDQCP1,1dDQCP2,1dDQCP3}.   We will comment more on this protection and related similar phenomena in other models in Sec. \ref{sec:generalization}. 

Given the importance of this symmetry let us now address whether the two massive phases -- apart from $\mathcal{T}$ symmetry breaking -- are topologically distinct in the SPT sense.  A simple way to  detect an  SPT phase is to identify a quantized topological response  to background gauge fields that couple to  the global symmetry\cite{Kapustin1,Kapustin2,akpsq}. Topological responses include the terms labeled by a continuous tuning parameter, such as the $\theta$-term, as well as terms labeled only by discrete parameters\cite{seiberg2013} (known as discrete theta terms).  As explained in Ref. \onlinecite{BS} in the presence of a non-trivial bundle of the background gauge fields, we require that 
\beq \label{sw_constraint}
w_2^{SO(3)_g}+w_2^{PSp(N_f)}=0 ~~ (mod~2).  
\eeq
Here $w_2$ refers to the second Steifel-Whitney class of the corresponding gauge bundle (the $SO(3)_g$ fields are dynamical, while the $PSp(N_f)$ fields are background), and we have restricted to flat spacetime manifolds\footnote{When the theory  is placed on a generic non-spin manifold we need to consider more general bundles that satisfy 
$w_2^{SO(3)_g}+w_2^{PSp(N_f)}+ w_2^{TM}= 0,~mod ~2$ where $TM$ is the tangent bundle of the manifold.}.  The notation $SO(3)_g$ here is due to the fact that when a topologically nontrivial background field for the global symmetry is turned on, the dynamical gauge field may be lifted to an $SU(2)/\Z_2 = SO(3)$ gauge field so long as \eqref{sw_constraint} is obeyed, which ensures that the fermions can be parallel-transported self-consistently.

For $m<0$, where we have $\theta=0$ for the $SU(2)$ gauge field, we can also choose the regularization such that it is in a  trivial phase of the global symmetry\footnote{Eq. \ref{eq:L} without the dynamical $SU(2)$ gauge field has a global $SO(4N_f)$ symmetry. We can choose a regularization such that a background $SO(4N_f)$ gauge field has trivial response in the $m<0$ phase. As a result, the $m<0$ theory will have trivial response for all subgroups of $SO(4N_f)$, including $PSp(N_f)$. }.

For $m > 0$, $\mathcal{T}$ and $\mathcal{P}$ are broken\footnote{The reader may wonder if we could have spontaneously broken $\mathcal{T}$ without breaking $\mathcal{P}$ or vice versa. This however is ruled out for the following reason. In this massive phase the universal low energy physics is captured  by a Lorentz invariant field theory which satisfies the $\mathcal{CPT}$ theorem. However, as emphasized before, in this theory there is no distinct notion of $\C$ that is independent from the global symmetries already considered, and hence we really have a $\P\T$ theorem. Then broken $\T$ implies broken $\P$ and vice versa, with a $\C$ transformation playing no role in the physics (see App. \ref{app:sym}
for details).}. Therefore we only need to examine whether the state is an SPT for the $\psp$ symmetry.
 
 For the group $\psp$ with $N_f$ odd, there is only one distinct type of topological response possible, namely the ordinary $\theta$ term\cite{seiberg2013}
\beq\label{ltop}
\mathcal{L}_{topo}=\frac{i\theta}{8\pi^2}\Tr(F\wedge F),
\eeq
where $F$ is the field strength of the $\psp$ background field.
The fact that there are no possible additional independent discrete terms arising from $\psp$ bundles that do not lift to $Sp(N_f)$ bundles is important for us, and is explained in App. \ref{app:instantons}. 
The quantization of the $\theta$-angle in \eqref{ltop} requires $\mathcal{T}$ or $\mathcal{P}$, which is absent on the $m>0$ side. Thus, there cannot be a nontrivial SPT for the $\psp$ symmetry. We conclude therefore that both of the two massive phases in the present theory are topologically trivial. \footnote{If there had been allowed discrete theta terms independent from Eq. \ref{ltop} then we could not make such a conclusion, since the coefficients of these terms are quantized even in the absence of $\T$, and hence can not be tuned away after the spontaneous breaking of $\T$. In principle, such a discrete term could arise when performing the path integral over the dynamical $SO(3)$ gauge field, since the constraint Eq. \ref{sw_constraint} means that the dynamical gauge field carries information about the discrete class $w_2^\psp$ of the background fields.}

A more physical  way to identify SPT phases is to gauge the global symmetry and study the quantum statistics and symmetry quantum numbers of the excitations. Therefore, let us gauge the $\psp$ symmetry and inspect the properties of its magnetic monopoles\footnote{Since $\pi_1(PSp(N_f))=\mathbb{Z}_2$, the $PSp(N_f)$ gauge group has a $\mathbb{Z}_2$ monopole.}. For the $m<0$ side, the monopoles are trivial bosons. However, on the $m>0$ side, the monopoles can potentially trap fermion zero modes. The diagnosis of a possible SPT phase can be done by asking whether there exists a monopole with trivial quantum numbers once the discrete symmetries are broken. Let us look at the simplest case with $N_f=1$, where the continuous global symmetry is $PSp(1)=SO(3)$. The system has two Dirac fermions, ($\psi_{\uparrow}$, $\psi_{\downarrow}$), which form an $SU(2)$ gauge doublet. The overall phase rotation of the $\psi$'s is the $z$-direction rotation of the $SO(3)$ global symmetry, with the other two generators involving rotations in particle-hole space.

Physically, the constraint \eqref{sw_constraint} means that a monopole configuration of the background $SO(3)$ gauge field is correlated with the flux configuration of the dynamical gauge field. Namely, a $\pi$-flux through the $z$-direction of the background field is associated with a $\pi$ flux through the $z$-direction of the dynamical gauge fields. Such a flux configuration explicitly breaks the $SO(3)$ global symmetry down to $SO(2)$. $\psi_\uparrow$ sees a $2\pi$-flux of this remaining $SO(2)$ while $\psi_\downarrow$ sees no flux, and therefore the monopole traps one fermion zero mode. We thus have two monopole states $\mathcal{M}^\dagger|0\rangle$ and $f^\dagger\mathcal{M}^\dagger|0\rangle$. These two states will be degenerate and have half-quantized $SO(2)$ charge if $\mathcal{T}$ symmetry is unbroken. However, these non-trivial properties are gone once $\mathcal{T}$ is broken. Physically, the fermion mode will no longer sit at zero energy and the degeneracy is violated, making the monopole trivial. The cases with odd $N_f>1$ have similar structure, which we do not list here.

Summarizing, we have shown that neither the $m < 0$ nor the $m > 0$ phase has any SPT order. Thus the $m < 0$ phase is a completely trivial gapped phase of the underlying bosons, while the $m > 0$ phase breaks  $\mathcal{T}$ symmetry but is otherwise trivial. The corresponding phase transition is certainly within the purview of standard LGWF theory in terms of a fluctuating order parameter field $\phi$ that is odd under $\mathcal{T}$ but is a singlet under $\psp$.  This leads to the  $3+1$-D Ising universality class, {\em i.e} the usual Gaussian fixed point with a marginally irrelevant $\phi^4$ interaction. However, in our model, the transition happens through a different fixed point which has emergent deconfined fermions and gauge fields. This is an example of a deconfined quantum critical point for a Landau-allowed transition in $3+1$-D.  Note that the emergent massless fermions transform projectively under $\psp$. Thus, in sharp contrast to the usual LGWF theory, gauge-invariant operators that transform nontrivially under $\psp$, \textit{e.g.} fermion bilinears in a singlet under the gauge $SU(2)$ but a nontrivial representation of $\psp$, will have power law correlations at this deconfined quantum critical point. This observation also implies that there is no possibility of describing this deconfined critical fixed point in any theory that involves just the fluctuating order parameter field $\phi$.

In a lattice realization, both the Landau and non-Landau allowed universality classes can presumably be accessed by tuning parameters (A schematic phase diagram is shown in Fig. \ref{fig:sPD}).

\subsection{Crossovers and critical exponents}
\label{xover}

Having established the zero temperature phase diagram, let us move on to the finite temperature phase diagram/crossovers and critical exponents for this LBL transition.  Near this fixed point the gauge coupling $g^2$ is dangerously irrelevant.  As usual this leads to two diverging length/time scales. First there is an obvious  length scale $\xi \sim \frac{1}{m}$. At shorter length scales (but still long compared to any lattice spacing) the system can be described in terms of massless fermions and gluons. At the length $1/m$ the gauge coupling $g^2$ is small
 and starts growing at longer distances. Confinement sets in only at a much larger length scale $\xi_{conf} \sim \xi^y$. The universal exponent $y$ is found by matching the RG flow of the massless theory with that of the pure gauge theory. To be explicit consider the well known RG flow  for non-abelian gauge theories with gauge group $G_g$ with $N_f$ fundamental massless fermions as a function of a dimensionless scaling parameter $l$: 
 \beq
 \label{RGflow}
 \frac{dg^2}{dl} = \left(\frac{11}{3}C_2 - \frac{4}{3} t N_f \right) \frac{g^4}{8\pi^2}
 \eeq
 Here $C_2$ is the quadratic Casimir of the gauge group $G_g$ and $t$ is defined by $\Tr(T^{a}T^{b})=t\delta^{ab}$  for generators $T^{a,b}$ of the Lie algebra in the fundamental representation.  Solving this equation for $N_f > \frac{11 C_2}{4t}$,  as assumed, at large RG scale $l$, we have 
 \beq
 \label{gsqlo}
 g^2(l) \approx \frac{1}{l}\frac{8\pi^2}{\left( \frac{4}{3} t N_f - \frac{11}{3}C_2 \right)}
 \eeq
 When the fermions have a bare mass $m$, we stop this RG flow at a scale $l_0$ at which $m e^{l_0} = \Lambda$ (the cut-off scale). 
 At longer scales $l \gg l_0$, the coupling grows as per the flow of the pure gauge theory, {\em i.e} Eq. \ref{RGflow} with $N_f = 0$. This leads to a solution
 \beq
 g^2(l \gg l_0) = \frac{g^2(l_0)}{1 - \frac{11 C_2}{24\pi^2}(l - l_0) g^2(l_0)}
 \eeq
 The coupling becomes strong at a scale $l^*$ (at which confinement occurs) such that 
 \beq
 l^* - l_0 = \frac{24\pi^2}{11 C_2 g^2(l_0)}
 \eeq
 Using Eq. 
 \ref{gsqlo}, we thus find 
 \beq
 l^* - l_0 = l_0 \left(\frac{4 t N_f - 11C_2}{11 C_2}\right).
 \eeq
 The ratio of the confinement scale and $\xi$ is $e^{l^* - l_0}$. It follows that the confinement length scale satisfies
 \beq
 \frac{\xi_{conf}}{\xi} = e^{l_0 \left(\frac{4t N_f}{11C_2} - 1\right)}
 \eeq
 Finally using $\xi = \frac{1}{m} = \frac{e^{l_0}}{\Lambda}$, we get $\xi_{conf} = \frac{1}{m^y}$ with 
 $y = \frac{4t N_f}{11C_2} > 1$. 
 Specializing to  $SU(2)$ gauge theory with $N_f$ fundamental fermions we have $y = \frac{N_f}{11}$.  At intermediate length scales between $\xi$ and $\xi_{conf}$ the system may be described in terms of massive deconfined fermions coupled to weakly interacting massless gluons.   
 
 \begin{figure}
\begin{center}
\includegraphics[width=0.9\linewidth]{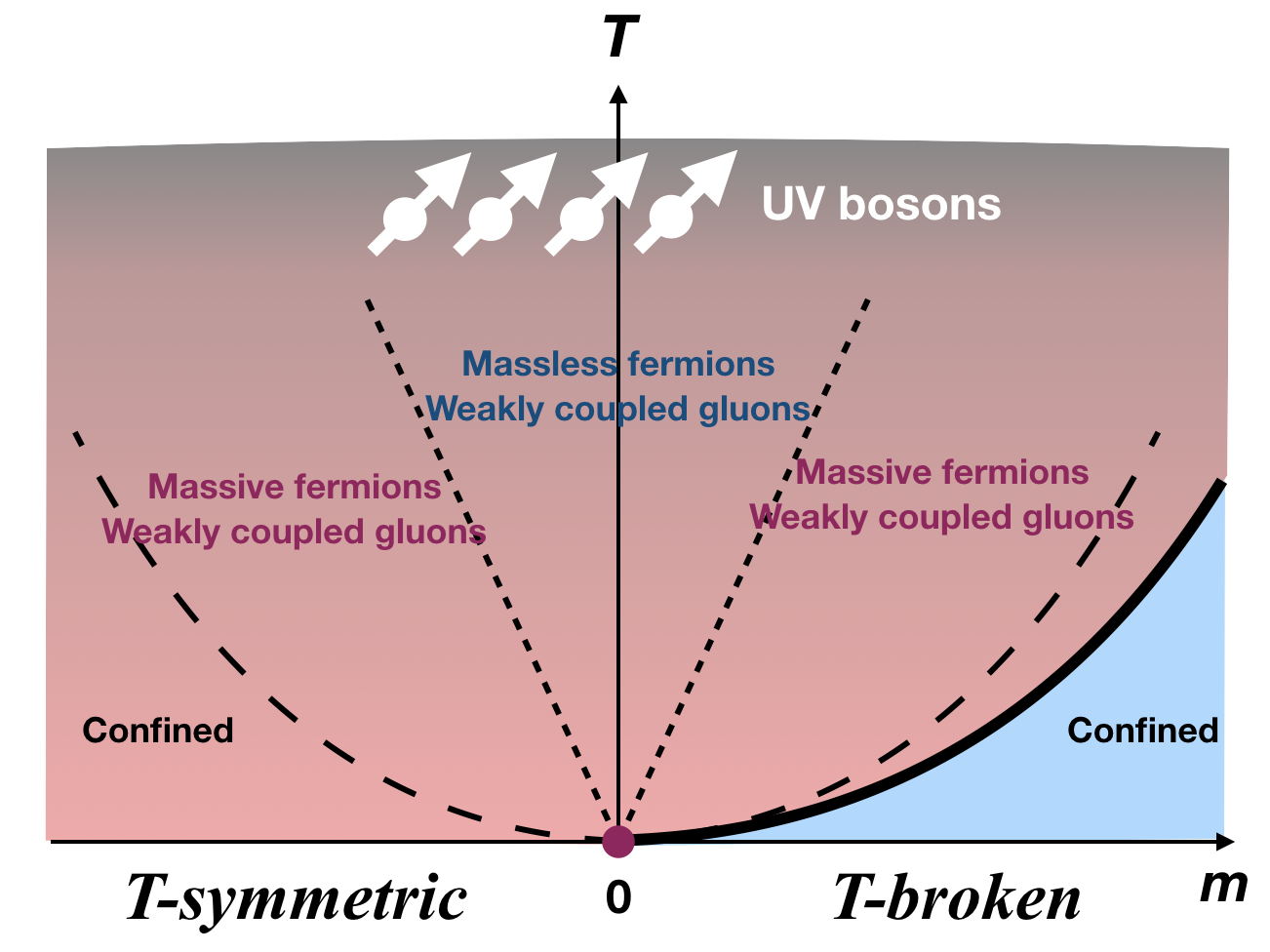}
\caption{A finite temperature phase diagram of the beyond Landau transition.}
\label{fig:PDT}
\end{center}
\end{figure}

It is important that the true nature of the two phases is only established beyond the confinement scale $\xi_{conf}$. In particular the time reversal breaking for $m > 0$ sets in only at this scale, which also sets the scale of the energy gaps in either phase.  These two distinct length scales will also lead to two distinct temperature scales for the crossovers at non-zero temperature (see Fig. \ref{fig:PDT}).  The broken time reversal symmetry will lead to a finite temperature phase transition with a $T_c \sim m^y$, {\em i.e} at the confinement scale.  Note that this transition is expected to be in the $3D$ Ising universality class though the zero temperature quantum phase transition is ``beyond Landau". 

It is interesting to obtain the critical exponents associated with this LBL universality class. The two length scales discussed above lead to two distinct correlation length exponents $\nu = 1$ (associated with $\xi$) and $\nu_{conf} = y$ (associated with $\xi_{conf}$).  The order parameter $\phi$ is a scaling field at the LBL fixed point. It may be identified with the operator $\bar{\psi} \gamma^5 \psi$, which is odd under $\mathcal{T}$, but is a singlet under $\psp$. As at the fixed point  the fermions are free, we see that $\phi$ has scaling dimension $\Delta_\phi = 3$, and thus its 2-point correlator decays as $\sim 1/x^6$ for space-time separation $x$. This corresponds to a very large value for the $\eta$-exponent: $\eta = 4$. This large $\eta$ is another striking contrast with the standard Ising universality class (where $\eta = 0$). 

Finally let us consider the exponent $\beta$ that describes how the order parameter rises as a function of the tuning parameter $m$ close to the transition.  As the broken symmetry only sets in at scale $\xi_{conf}$, we have $\beta = \nu_{conf} \Delta_\phi = 3y$. This leads to a very slow rise of the order parameter compared to the usual Ising universality class (where $\beta = 1/2$). The exponents are summarized in Table. \ref{tab:exponents}.

\begin{table}[]
    \centering
    \begin{tabular}{|c|c|c|}
    \hline
        exponents & Ising & DQCP\\ \hline
        $\nu$ & $1/2$ &  $\nu_1=1$,\ $\nu_2=y=\frac{4tN_f}{11C_2}>1$ \\ \hline
        $\eta$ & $0$ & $4$ \\ \hline
        $\beta$ & $1/2$ & $3y$ \\ \hline
    \end{tabular}
    \caption{Comparison of critical exponents. There are two entries for $\nu$ at the DQCP, as there are two diverging length scales.}
    \label{tab:exponents}
\end{table}


\section{Pure gauge theories at $\theta = \pi$} 
Given its crucial importance we now review what is known about the pertinent physics of pure non-abelian gauge theories with a theta term at $\theta = \pi$ in $3+1$-D.  
The appealing arguments for the $\mathcal{T}$  broken confining phase are two-fold. The early evidence\cite{witten1980large,Witten} is by inspecting the ground state energy, $E(\theta)$, as function of $\theta$-angle (for $SU(N_c)$ pure Yang-Mills gauge theory). The compatibility between large-$N_c$ scaling and the periodicity of the $\theta$-angle implies that the ground state energy $E(\theta)$ must have multiple branches and level crossings at $\theta=\pi$ that break the $\mathcal{T}$ symmetry. This is shown explicitly through holographic methods for some large-$N_c$ theories\cite{Witten}. The same argument is readily seen to also apply to $Sp(N_c)$ gauge theories in the large $N_c$ limit which we will consider below.  What happens at smaller  $N_c$ (in particular for $SU(2)$ gauge theory) is left unresolved by this reasoning.

Recently a new anomaly argument has provided strong constraints on the IR fate of such theories.  Specifically, it precludes a trivial confining phase for pure $SU(N_c)$ gauge theory with $\theta=\pi$\cite{zohar2017,Gaiotto2018,WangQCD,WangAnomaly,GuoWang}.  Pure $SU(N_c)$ gauge theory  has unbreakable electric strings in the center $\Z_{N_c}$ of the gauge group. This is captured by saying that the pure gauge theory has a $\Z_{N_c}$ 1-form symmetry\cite{1form}. At $\theta = \pi$ the theory is also time reversal symmetric. However it is shown in Ref. \onlinecite{zohar2017} that for any $N_c\in 2\mathbb{Z}$, the time reversal symmetry and the $\Z_{N_c}$ 1-form symmetry have a mixed anomaly. Physically, one can show that the time reversal symmetry will be broken if one gauges the 1-form symmetry, which is a signature for a mixed anomaly. The result of gauging the $\Z_{N_c}$  1-form symmetry of the pure $SU(N_c)$ gauge theory is an $SU(N_c)/\Z_{N_c}$ gauge theory\cite{seiberg2013}. While the $\theta$-angle is still $\pi$, it has a different periodicity in the gauged theory, namely $2\pi N_c$\cite{seiberg2013,zohar2017}. Therefore, the time reversal symmetry $\mathcal{T}$ (as well as $\mathcal{CT}$) is explicitly broken, since $-\pi$ and $\pi$ are not equivalent $\theta$-angles in the gauged theory. 

This mixed anomaly is an indication that the system cannot have a gapped featureless ground state. Consider any completely gapped ground state which is also confining. Then the $\Z_{N_c}$ 1-form symmetry is unbroken.  The mixed anomaly then suggests that $\mathcal{T}$ is broken.  The alternate possibility that the theory saturates the mixed anomaly in a gapped state by developing topological order ({\em i.e} through a Topological Quantum Field Theory) is  forbidden\cite{ccko2019}: A theory with this mixed anomaly  is ``symmetry-enforced gapless" in the sense discussed for ordinary 0-form symmetries in Ref. \onlinecite{senthilDIII}. The conclusion therefore is that a gapped ground state that is also confined necessarily breaks $\mathcal{T}$ symmetry for $SU(N_c)$ gauge theory for any even $N_c$ at $\theta = \pi$. Alternately, the anomaly may also be saturated by a gapless state that preserves both the 1-form and $\mathcal{T}$ symmetries\footnote{A proposal for such a gapless symmetric state is still lacking.}. 

For pure $Sp(N_c)$ gauge theories, there is a $\Z_2$ 1-form symmetry for any $N_c$ associated with the $\Z_2$ center of the gauge group.  At $\theta = 0, \pi$, the theory is $\mathcal{T}$ symmetric. Again at $\theta = \pi$ there is a mixed anomaly between the $\Z_2$ 1-form symmetry  and $\mathcal{T}$ so long as $N_c \neq 0~mod~4$ (see App. \ref{app:instantons} for a brief review and references). For these values of $N_c$ (which we will restrict to) the rest of the discussion is identical to that in the previous paragraph. 

Although the anomaly argument cannot completely tell us the IR dynamics of these pure gauge theories, a time reversal broken confined phase is the simplest possible outcome which is consistent with all the constraints.  If we wish to be really safe, we can restrict to large-$N_c$ gauge theories, in which the statement of $\mathcal{T}$ breaking in the $\theta=\pi$ vacuum is controlled. 

\section{Large-$N_c$ generalization\label{sec:generalization}}
\subsection{$Sp(N_c)$ generalization}
It is easy to generalize the parton theory to different gauge groups without changing the global symmetry of the system. We use the construction from Ref. \onlinecite{BS}. Formally, we consider the same Lagrangian in Eq. \ref{eq:L} with   $Sp(N_c)$ gauge fields  and fermions in the $Sp(N_c)$ fundamental representation. The global symmetry of this theory is exactly the same as the $SU(2)$ theory, which can be thought of as a special case of $N_c=1$ in this series.  We always consider $N_f$ large enough that, at the massless point, the theory is infra-red free.  For finite $N_c$ we also restrict to $N_c \neq 0, ~mod~4$ for reasons explained in the previous section. Then with the further assumption (which is surely correct for large enough $N_c$) that the pure gauge theory at $\theta = \pi$  breaks $\mathcal{T}$ symmetry but is otherwise trivial, the  phase diagram of this series of parton theories are the same as the scenario discussed above for the $SU(2)$ case.   Therefore, these theories in the large-$N_c$ limit serve as an exactly soluble limit where this phase diagram can be established with confidence. 

The crossover exponent $y$ discussed in the Sec. \ref{xover} will depend on $N_c$ and $N_f$ in a manner easily computable from the known RG flows of these theories. The remaining part of that discussion will be unchanged.

\subsection{$SU(N_c)$ generalization}

We can also generalize to the case where the gauge group is $SU(N_c)$, with $N_c>3$. For all even $N_c$, in the same sense as in previous sections, this theory again describes a UV system of  bosons. The global internal symmetry of the theory is $G = [U(N_f)/\Z_{N_c}] \rtimes \Z_2^\C$, where the quotient by $\Z_{N_c}$ comes from the action of the gauge group.\footnote{The essential difference between the $SU(2)$ and the $SU(N_c>2)$ cases is that the fundamental representation of $SU(2)$ is pseudoreal (isomorphic to its complex conjugate), while for $N_c>2$ it is complex. Hence a distinct action of charge conjugation may be defined when $N_c>2$, while for $N_c=2$ its action is "absorbed" into the flavor symmetry, enlarging it to $Sp(N_f)$. } When $N_f$ is odd, integrating out the fermions at $m>0$ again produces theta terms at $\theta=\pi$ for the $SU(N_c)$ gauge fields and for the $U(N_f)/\Z_{N_c}$ background fields, if they are present
(again, we regularize so that the topological response is trivial for $m<0$).

As mentioned above, one expects that the $SU(N_c)$ gauge theory at $\theta=\pi$ spontaneously breaks $\T$, at least at large $N_c$.
The differences with respect to the $SU(2)$ theory come up when one addresses the question of whether or not the $m>0$ and $m<0$ phases can differ by an SPT of the $U(N_f)/\Z_{N_c}$ symmetry. Again, after time reversal is broken, the only possible SPT that could remain would be one with a discrete topological response, which cannot be re-expressed 
in terms of a continuous theta term. While the details depend on the exact choices of $N_c$ and $N_f$, we show in App. \ref{app:instantons} that when ${\rm gcd}(N_c,N_f)=1$ (this includes the cases of interest, namely odd $N_f$ and possibly large $N_c$), no discrete topological responses for the $U(N_f)/\Z_{N_c}$ symmetry exist. Therefore we again have an example of an exotic phase transition between a trivial gapped phase and a phase with trivially broken $\T$ symmetry.\footnote{Technically, we should also examine the possibility of SPT phases protected by the unbroken $\C$ symmetry. However, no such SPT phase exists. $\C$ is a $\Z_2$ unitary symmetry, and it is known that there is no SPT for bosons in $3+1$-D protected purely by such a symmetry. This only leaves the possibility that there may be an SPT protected by the combination of $U(N_f)/\Z_{N_c}$ and $\mathcal{C}$. However, the existence of such an SPT would require a discrete theta term that couples a background $\Z_2$ gauge field for $\mathcal{C}$ to a field coming from the $U(N_f)/\Z_{N_c}$ bundle. Such a term does not exist when $\gcd(N_f,N_c)=1$, since then the $U(N_f)/\Z_{N_c}$ bundle has no discrete classes which the $\C$ gauge field can couple to.} 

\section{Alternate scenarios for $SU(2)$ gauge theory} 
We now return to the $SU(2)$ gauge theory and consider the possibility that the pure gauge theory at $\theta = \pi$ is in a phase distinct from the simple $\mathcal{T}$ and $\mathcal{P}$ broken phase assumed so far. As specific examples we consider  two distinct and interesting possibilities that are consistent with the mixed anomalies.  In the full gauge theory ({\em i.e} including the matter fields) these will be interpreted as quantum spin liquid phases of the underlying bosons. In both these cases we show that the massless point describes an extremely novel phase transition between these spin liquids and a trivial gapped phase. 

\subsection{Confinement transition of a $U(1)$ spin liquid}
First consider the possibility that the infrared fate of the $SU(2)$ gauge theory with $\theta=\pi$ is a deconfined $U(1)$ spin liquid with unbroken $\T$ symmetry. Such a phase breaks the global $\Z_2$ 1-form symmetry spontaneously.  The properties of such a phase can be readily accessed by modifying the gauge theory by including a coupling to a Higgs field in the adjoint ({\em i.e} spin-1) representation of the $SU(2)$ gauge group.  Introducing such a Higgs field retains both the $\Z_2$ $1$-form symmetry and $\mathcal{T}$, as well as their mixed anomaly. Condensing this Higgs field then leaves behind a residual unbroken $U(1)$ gauge group and therefore leads to a deconfined $U(1)$ gauge theory.

Assuming this is the fate of the pure gauge theory at $\theta = \pi$, let us now describe the physics of the full system that includes the massive fermionic matter fields. The fermions are charged under the unbroken $U(1)$ gauge group. In addition there will be gapped magnetic monopoles whose properties we describe below. In terms of the UV bosons, the resulting state is a $U(1)$ quantum spin liquid enriched by the global $PSp(N_f) \times \mathcal{T}$ symmetry. 

 Due to the $\theta=N_f\pi$ in the original $SU(2)$ gauge theory, this  $U(1)$ gauge theory inherits a $\theta$-angle at $\theta_{U(1)}=2N_f\pi$. The Lagrangian for the $U(1)$ spin liquid state can be written as the following, 
\bea
\nonumber
\mathcal{L}_{U(1)}&=&\sum_{i=1}^{N_f}\sum_{\sigma=\uparrow,\downarrow}i\bar{\psi}_{i,\sigma}(\gamma_\mu(\partial_\mu-(-)^\sigma ia_\mu)-m)\psi_{i,\sigma}\\
&&+\frac{1}{4e^2}f^2+\frac{2N_f\pi}{8\pi^2}f\wedge f,
\label{eq:u1}
\eea
where we make a gauge choice for the residual $U(1)$ gauge field to be the $S_z$ component of the $SU(2)$ gauge field from the UV. At low energies (below  both the fermion mass and the monopole mass), the physics is that of free Maxwell theory which has both an electric and magnetic $U(1)$ one-form symmetry (which are both spontaneously broken).  The $SU(2)$ gauge theory in Eq. \ref{eq:L} has an emergent $\Z_2$ 1-form symmetry at energy much lower than the fermion mass which  maps - at low energy- to the $\Z_2$ subgroup of the emergent (but spontaneously broken) electric 1-form symmetry in the $U(1)$ gauge theory. 

Let us consider the global symmetry properties of the $U(1)$ spin liquid state. As a warm-up we first consider the case of $N_f=1$ where the global symmetry is $G=PSp(1)\times \T=SO(3)\times \T$. We can do a particle-hole transformation on the $\psi_{\downarrow}$ fermions by defining 
$(f_{\uparrow},f_{\downarrow})=(\psi_{\uparrow},-\psi^\dagger_{\downarrow})$. The $f$-fermions carry the same charge under the $U(1)$ gauge field. The action of the $SO(3)$ symmetry is then manifest,
with the vector $(f_\upa,f_\doa)^T$ transforming projectively under the left action of $SO(3)$ as a spinor. The $f$-fermions are the electric charge excitations of the $U(1)$ gauge theory, 
and their time reversal transformation  is the same as in Eq. \ref{eq:T}.
Thus in our convention the electric charge is time reversal odd in this $U(1)$ gauge theory (and correspondingly the magnetic charge will be time reversal even). 

Next, we consider the properties of the magnetic monopoles in this $U(1)$ theory. Due to the non-zero $\theta$-angle, the monopole can carry nontrivial quantum numbers under the global symmetry. Let us put the system on a large sphere. The surface of Eq. \ref{eq:u1} hosts two massless Dirac fermions coupled to the dynamical $U(1)$ gauge field. Consider a configuration of $2\pi$ magnetic flux of the $U(1)$ gauge field coming out of the bulk. This monopole configuration will trap two fermion zero modes, which we label as $\eta_\uparrow$ and $\eta_\downarrow$. There are in total four states, labeled by  $\mathcal{M}^\dagger|0\rangle$, $\eta_\uparrow^\dagger \mathcal{M}^\dagger|0\rangle$, $\eta_\downarrow^\dagger \mathcal{M}^\dagger|0\rangle$,
and $\eta_\uparrow^\dagger\eta_\downarrow^\dagger \mathcal{M}^\dagger|0\rangle$. Two of the four states, $\{\eta_\uparrow^\dagger \mathcal{M}^\dagger|0\rangle, \eta_\downarrow^\dagger \mathcal{M}^\dagger|0\rangle\}$, are gauge neutral, and transform as a spin-$1/2$ under the global $SO(3)$ symmetry. Notice that the $\T$ symmetry flips the gauge charge. Therefore, the monopole configuration is time reversal invariant. The two states $\{\eta_\uparrow^\dagger \mathcal{M}^\dagger|0\rangle, \eta_\downarrow^\dagger \mathcal{M}^\dagger|0\rangle\}$ form a Kramers doublet under time reversal symmetry. To summarize, the $U(1)$ spin liquid has electric charge that is a spin-$1/2$ fermion and magnetic monopole that is a spin-$1/2$ Kramers doublet boson. We denote this phase as $E_{f\frac{1}{2}}M_{\frac{1}{2}T}$.

We can now generalize the analysis to all odd $N_f>1$. Then the  electric charges of the $U(1)$ spin liquid are fermions that carry the projective representation of the global   $PSp(N_f)$ symmetry. The magnetic monopoles also carry the projective representation of the $PSp(N_f)$ and transform under time reversal with $\T^2=-1$. We can see this from the following: first, consider the same monopole configuration above. In this case, it will trap $2N_f$ fermion zero modes, labeled by $\eta_{i}$, $i=1,2,...,2N_f$. To construct the gauge neutral monopole, we need to consider states that half fill the zero modes, namely $N_f$ out of the $2N_f$ zero modes. These states can be labeled by $|T_{\{i_1,i_2,...,i_{N_f}\}}\rangle=\eta_{i_1}^\dagger\eta_{i_2}^\dagger...\eta_{i_{N_f}}^\dagger\mathcal{M}^\dagger|0\rangle$, where $i_1,...,i_{N_f}=1,2,...,2N_f$. The $\Z_2$ center of $Sp(N_f)$ group  acts on the $\eta_i$ modes as $\eta_i\mapsto-\eta_i$. Since $N_f$ is odd, the $|T_{\{i\}}\rangle$ states are odd under the center symmetry as well. Therefore, they transform in a projective representation of the $PSp(N_f)$ symmetry. In addition, the pairs of states $|T_{\{i\}}\rangle$ and $|T_{\{j\}}\rangle$ with non-overlapping sets $\{i\}$ and $\{j\}$ form Kramers doublets under the $\T$ symmetry. In terms of the symmetry realization on the $E$ and $M$ particles this $U(1)$ spin liquid can thus also be denoted\footnote{There is however one further subtlety in fully characterizing it this way: there may be an additional SPT phase protected by the global symmetry.  See  Refs. \onlinecite{TRu1,ZouU(1)}.} as $E_{f\frac{1}{2}}M_{\frac{1}{2}T}$.    

The $U(1)$ spin liquid state we described above is not anomalous\cite{ZouU(1)}. Within the assumption made in this subsection, it admits a direct continuous phase transition, described by the massless $SU(2)$ gauge theory,  to a trivial gapped confined phase.
Such a second order confimenent transition   is hard to describe in any simple way in terms of the obvious excitations of the $U(1)$ spin liquid phase. The naive route of condensing the bosonic $M$ particle will not lead to a symmetric confined phase due to its non-trivial symmetry properties (and besides, at weak coupling, a theory of $M$-condensation is believed to be first order so long as $N_f$ is not too large). 
 The $SU(2)$ gauge theory provides an extremely novel new path for the evolution from the $U(1)$ spin liquid to a trivial confined phase with all the global symmetry preserved. As the fermionic charge in the $U(1)$ spin liquid becomes light,
 the $U(1)$ gauge theory gets enlarged to an $SU(2)$ gauge theory. The critical point is the $SU(2)$ gauge theory with massless fermions. We notice a remarkable feature of the theory is that the mass gap for the monopole excitations of the $U(1)$ spin liquid also becomes light, $m_{M}\sim m_E^y$, as the mass of fermionic charge $m_E$ is reduced, which is potentially the reason for the transition to evade conventional expectations. After the fermion mass becomes negative, the $SU(2)$ gauge theory confines to a trivial state.

 \subsection{Confinement transition of a $\Z_2$ spin liquid}
 Now we consider yet another possibility: that the infrared fate of the $SU(2)$ gauge theory with $\theta = \pi$ is a deconfined $\Z_2$ spin liquid with unbroken $\mathcal{T}$ symmetry. This phase also breaks the global $\Z_2$ 1-form symmetry but is gapped, unlike the $U(1)$ spin liquid considered above.  Let us directly consider the implications of this assumption for the full theory that includes the gapped fermionic matter field.  
 
 To infer the properties of such a $\Z_2$ spin liquid it is convenient to imagine reaching it from the $U(1)$ spin liquid of the previous subsection by condensing pairs of the fermionic $E$ particles:  clearly such a pair of fermions can be a $PSp(N_f)$ singlet boson that also transforms trivially under $\mathcal{T}$; condensing these then leads to a symmetry preserving $\Z_2$ spin liquid. There are gapped fermionic (``Bogoliubov") quasiparticles that carry the $\Z_2$ gauge charge and transform projectively under $PSp(N_f)$. In addition there are, as usual, tensionful $\Z_2$ flux lines around which the $\Z_2$ charges braid with $\pi$ phases. 
 
 If this is what happens for $m > 0$ in the $SU(2)$ gauge theory, then the massless point again represents a highly non-trivial quantum critical confinement transition to a trivial symmetric phase from such a $\Z_2$ spin liquid. We are not aware of any previous description of a continuous $T = 0$ confinement transition from a $\Z_2$ spin liquid with a fermionic $\Z_2$ gauge charge. 
 
 In conclusion we see that with any of the three scenarios - $\mathcal{T}$ broken but otherwise trivial, the $U(1)$ quantum spin liquid, or the $\Z_2$ quantum spin liquid - the massless $SU(2)$ gauge theory describes highly non-trivial quantum critical phenomena.

\section{Contrast with other critical phenomena}
It is interesting to contrast the behavior we have found for the Landau ordering transitions with other critical phenomena that, at first sight, might seem similar.

It is of course well known that even within the standard LGWF paradigm the predictions of Landau mean field theory are modified by fluctuations.  Of interest to us however are situations where the LGWF paradigm itself  is challenged at a Landau ordering transition. Here we discuss two  examples. The first  is the 3-state Potts model in $D = 2$ space-time dimensions. Here Landau mean field theory predicts a first order transition; however it is known that the transition can be second order with non-trivial critical exponents given by a CFT. The second example (which is closer to the ones in this paper) is the abelian Higgs model with $N$ complex scalars in $2+1$-dimensions. When $N$ is sufficiently large this model is known to have a second order transition described by the non-compact $CP^{N-1}$ universality class\footnote{This means that monopole operators in the $U(1)$ gauge field are all irrelevant at the IR fixed point for sufficiently large-$N$}. This is a Landau ordering phase transition between a trivial gapped phase and a phase with broken global symmetry but which is otherwise trivial. The gauge invariant order parameter for the phase transition is  a Hermitian matrix 
\beq
Q_{ab} = z_a^\dagger z_b - \frac{\delta_{ab}}{N}  z^\dagger z
\eeq
(here $z_a$, $a = 1,....,N$ are the complex scalar fields). 
An expansion of the free energy  in terms of the matrix  $Q$ admits a cubic term (for any $N > 2$) and hence, in Landau mean field, the transition is first order.  Nevertheless the transition is allowed to be second order. 

From the Landau mean field point of view, it may be surprising that both the 3-state Potts in 2D and the 3D large-$N$ abelian Higgs model may still fit in within the LGWF paradigm.  To see this concretely consider the LGWF Lagrangian for the fluctuations of the order parameter $Q$ for the 3D abelian Higgs model: 
\bea
\label{landauQ}
\nonumber
{\cal L}_Q &=& \Tr ((\partial Q)^2)  + r \Tr(Q^2) \\&&+ v \Tr(Q^3) + u \Tr(Q^4) + w \Tr(Q^2)^2 
\eea
The transition is accessed by tuning one parameter, for instance $r$. The critical theory is strongly coupled in the IR as $(v,u,w)$ are all strongly relevant at the Gaussian critical fixed point at $(r,v,u,w) = (0,0,0,0)$. Such a description has previously been considered in Ref. \onlinecite{AdamNLSM} which contains much useful discussion. Presumably for some range of $(v,u,w)$ within the critical manifold this theory flows to the critical fixed point easily accessed in the gauge theory description through the Lagrangian
\beq
\label{nccpN}
{\cal L}_z = |\left(\partial - ia \right) z|^2 + \tilde{r} |z|^2 + \tilde{u} |z|^4 + ....
\eeq
(The ellipses denote other allowed local terms, eg, a Maxwell term for the $U(1)$ gauge field $a$). 

While we do not know for sure that such an RG flow exists, it does not seem to violate any principle we are aware of. We thus conjecture that such a flow is indeed possible. Indeed a lattice version of Eq. \ref{landauQ} can be solved explicitly in the large-$N$ limit\cite{DIVECCHIA}, and yields criticality described by the universality class of Eq. \ref{nccpN}. Thus it is not unreasonable to expect that the continuum Eq. \ref{landauQ} (defined by perturbing the free theory by $(r,v,u,w)$) can also flow to the non-compact $CP^{N-1}$ fixed point. Accepting this we see that the critical fixed point of Eq. \ref{nccpN} can indeed be described - as a matter of principle - by the LGWF theory of Eq. \ref{landauQ}, albeit in a cumbersome and inconvenient manner. 

The exact same comments also apply to models studied in Ref. \onlinecite{subiretal2019} which involved Higgs transitions in $SU(2)$ gauge theories to describe Landau ordering transitions in $2+1$-D. Within naive Landau mean field this transition is first order, but the gauge theory gives a possible route to a second order transition through a deconfined critical point. The corresponding fixed point can presumably again be reached through an RG flow of the standard LGWF action.

In contrast in the examples described in this paper with $G \times \mathcal{T}$ symmetry, the LGWF formulation will be a theory solely in terms of the Ising order parameter field $\phi$ which is singlet under $G$.  Thus within this formulation operators that transform under $G$ will not be scaling fields, and will have exponential correlations.  However at the critical fixed points we have described, such operators are scaling fields and have power law correlations. Thus even as a matter of principle, the LGWF description cannot capture these critical fixed points. 

A different phenomenon is the role that the unbroken global symmetry $G$ plays in protecting the new universality class. A familiar example of a related phenomenon happens at the superfluid-insulator transition of bosons on a clean lattice\cite{sachdevbook}. Depending on whether the transition is tuned by interactions at fixed commensurate density or by chemical potential there are different universality classes.  From the point of view of the low energy LGWF theory the  former  universality class is protected by the presence of a particle-hole symmetry  at the critical fixed point.  Note however that the relevant operator that is forbidden by $C$ symmetry is $-\mu j^0$ where $j^0$ is the boson density. This operator has power law correlations through out the superfluid phase. This is a difference with the examples in this paper where operators that transform (or generate) the $G$ symmetry have exponentially decaying correlations everywhere except at the critical point. 

Recently a $1d$ generalization of the deconfined quantum critical points has been discussed in Refs. \onlinecite{1dDQCP1,1dDQCP2,1dDQCP3,Mudry1d}. Consider a $1d$ spin-$1/2$ chain with $\Z_2^x\times \Z_2^z$ spin rotation symmetry as well as translation symmetry. The low energy theory can be most conveniently presented in the language of a $O(4)$ nonlinear $\sigma$-model with a Wess-Zumino-Witten term and anisotropies. The components of the $O(4)$ vector $\bm{n}$ parameterize the four leading ordering tendencies ($x$-FM, $y$-AFM, $z$-FM, VBS) in the model. The $1d$ DQCP describes a continuous phase transition between a $z$-ferromagnetic state and a VBS ordered state, which break $\Z_2^x$ and translation, respectively. Similar to our LBL example, the $\Z_2^z$ symmetry, which flips the $x$-FM and $y$-AFM order, is not broken in either phase or at the phase transition. Nonetheless the presence of the $\Z_2^z$ symmetry is important. If one allows explicit $\Z_2^z$ breaking, one can add an additional relevant perturbation, namely $h S_x$, that can destroy the critical point and lead the system to a spin polarized phase. 

Finally, in App. \ref{app:1dSchw} we discuss a $1+1$-dimensional model of $N_f$ fermions coupled to a $U(1)$ gauge field (known as the Schwinger model). We show how, despite some superficial similarities, the phase transition in this model exhibits a different, less surprising phenomenon, from the one studied in this paper. Specifically this $1+1$-D model contains a non-trivial second order phase transition between a trivial gapped phase and a $\mathcal{P}$-broken phase which both preserve a continuous global symmetry of the model. However, the symmetry broken ground states also differ from the trivial state by having a non-trivial SPT response to background gauge fields for the unbroken symmetry. 

\section{Discussion} 
The examples discussed in this paper concretely show how order parameter fluctuations might distract from the true critical behavior at some Landau ordering transitions even, in the ``standard" case where there is no other topological or other exotic order in either phase, and the transition is Landau-allowed.  In these examples the essential transition is actually a  topological phase transition of emergent fermions which, however, disappear from view at the longest scales in either phase due to confinement. 

It is natural to wonder if in $3+1$-D there are continuous Landau-forbidden quantum phase transitions between phases that themselves are Landau allowed. Such Landau forbidden transitions are well known in $2+1$-D but there are no examples in $3+1$-D that are known to us. In App. \ref{app:3dsu2b} we show that a bosonic version of the theories discussed above gives a natural construction of deconfined critical fixed points for such Landau-forbidden transitions in $3+1$-D, subject to the caveat that we work with a (sometimes rather fancifully) high number of bosons.

While the specific models in which we have been able to demonstrate these phenomena may seem esoteric from a condensed matter point of view,  they are concrete and hence settle matter of principle questions. We hope that future work finds similar phenomena in simpler models. Perhaps more importantly the examples discussed here lend some moral support to phenomenological ideas in heavy fermion and other experimental systems that invoke physics beyond order parameter fluctuations (in a metallic environment). 

\textit{Note}: After the completion of this work, we notice Ref. \onlinecite{WangU1} appeared, which had partial overlap with our work and focused more on the discussion of scenario involving U(1) spin liquid.

\section*{Acknowledgement} We thank Clay Cordova, Dominic Else, Shenghan Jiang, Zohar Komargodski, Max Metlitski, Olexei Motrunich, Adam Nahum, Nathan Seiberg, Ryan Thorngren, Cenke Xu, and Yi-zhuang You  for inspiring discussions, and Ho Tat Lam and Yuji Tachikawa for helpful correspondence. ZB is supported by the Pappalardo fellowship at MIT. ZB is also partially supported by KITP program on topological quantum matter by the National Science Foundation under Grant No. NSF PHY-1748958. EL is supported by the Fannie and John Hertz Foundation and the NDSEG fellowship. TS was supported by NSF grant
DMR-1911666, and partially through a Simons Investigator Award from the Simons Foundation. This work was also supported by the Simons Collaboration on Ultra-Quantum Matter, which is a grant from the Simons Foundation (651440, TS).

\appendix

\section{Global symmetries}
\label{app:sym}

In this appendix, we discuss the global symmetries present in the $SU(2)$ gauge theory. As in Ref. \onlinecite{BS}, it is helpful to start from a theory of $4N_f$ Majorana fermions. In the free theory with just Majoranas, the internal symmetry group of the theory is $O(4N_f)$. We will choose $(-,+,+,+)$ signature with $\gamma$ matrices $\gamma_0 = -i\mu^y \tp \sigma^x, \, \gamma_1 = \mu^y\tp \sigma^y, \, \gamma_2 = \mu^x \tp \unit, \, \gamma_3 = \mu^z\tp \unit, \, \gamma_5 = \mu^y \tp \sigma^z$, where $\mu^i$ and $\sigma^i$ are Pauli matrices. 
This choice is particularly convenient since all the $\gamma_\mu$ matrices are real, and so the fields manifestly transform in a real representation of the
global symmetry group. 

To study the $SU(2)$ gauge theory, we re-package the fermions by defining the fields 
\be X_v = \frac{1}{\sqrt{2}}(\chi_{\up v} \sigma^0 + i \eta_{\down v} \sigma^x + i \chi_{\down v} \sigma^y + i \eta_{\up v} \sigma^z). \ee
Here the flavor index is $v \in \{1,\dots,N_f\}$, and the notation is such that the Majoranas are grouped into complex fermions of $SU(2)$ spin $\sigma$ and flavor $v$ via $\psi_{\sigma v} = \chi_{\sigma v} + i\eta_{\sigma v}$. In this notation, the free kinetic term is $i\sum_v\Tr[\bar X_v \slashed{\p} X_v]$. 

We then gauge the right $SU(2)$ action on $X$. After gauging, the internal global symmetry that remains is $N_{O(4N_f)}(SU(2)) / SU(2)$, where the normalizer group $N_{O(4N_f)}(SU(2))$ contains all those elements $R \in O(4N_f)$ such that for any $V\in SU(2)$, we have $R^{-1} V R \in SU(2)$. 
Naively, this leaves a $U(2N_f)$ symmetry acting on $X$ on the left, since the left action commutes with the $SU(2)$ action by construction. 
However, not all matrices in $U(2N_f)$ are allowed: the reality condition of the Majoranas means that in fact the matrices acting on the left must be in $Sp(N_f)$. This can be shown by requiring that the map $X \mt (\unit_{N_f} \tp J)^{-1} X J$ with $J \equiv -i\sigma^y$ is equivalent to complex conjugation on the Dirac fields $\psi_{\sigma v}$ both before and after the left action: any unitary $U$ acting on $X$ on the left must satisfy $U^T(\unit_{N_f} \tp J)U = (\unit_{N_f} \tp J)$, implying $U\in Sp(N_f)$. 
Since the element $-\unit\in Sp(N_f)$ acts on the fields in the same way as the element $-\unit \in SU(2)$, we actually only get a $PSp(N_f) \equiv Sp(N_f) / \Z_2$ global symmetry, and one can show that $PSp(N_f)$ is in fact the entire remaining internal symmetry group. \footnote{Indeed, $PSp(N_f) \times SU(2)$ is a maximal subgroup of $SO(4N_f)$, meaning that it is not a subgroup of any proper subgroup. 
Since the normalizer of a subgroup is itself a subgroup, the normalizer cannot possibly be any bigger than $PSp(N_f) \times SU(2)$, and so in fact must be equal 
to it---quotienting by $SU(2)$, we indeed obtain $PSp(N_f)$ as the full internal symmetry group.} 

Now we discuss the discrete symmetries of time reversal and parity. We will take $\T$ to be an anti-linear operator which includes an action of complex conjugation on the dynamical fields (our definition of $\T$ corresponds to what is often called $\C\T$ in the literature).
One can check that the transformation 
\be\ba  \T : X(t,\bfx) &\mt \gamma_0\gamma_5 X^*(-t,\bfx)\\ & = \gamma_0\gamma_5 (\unit_{N_f} \tp J)^{-1}X(-t,\bfx) J,\ea\ee 
leaves the action invariant provided that the $SU(2)$ gauge field transforms as
\be \ba \T  : a^I_0(t,\bfx) &\mt - a^I_0(-t,\bfx), \\ 
 a^I_i(t,\bfx)
 &\mt a^I_i(-t,\bfx),\ea\ee
 so that the gauge coupling $\sum_v\Tr[\bar X_v \slashed{a}^I X_v \sigma^I/2]$ is left invariant, with $I=1,2,3$ the gauge index. 
The reason why we choose to call this transformation $\T$ instead of $\CT$ is because it only acts on the Lorentz indices of the fields in the theory---if we had not included the complex conjugation of the fermion fields, $\T$ would be required to act on the internal gauge indices of the fields as well.\footnote{Indeed, define an antilinear operation $\CT : X(t,\bfx) \mt \gamma_0\gamma_5X(-t,\bfx)$ on the fermion fields by taking the action of $\T$, but omitting the complex conjugation on $X$. $\CT$ invariance of the gauge coupling is achieved provided that the $SU(2)$ gauge field transforms as $\CT : a_\mu(t,\bfx) \mt (-1)^{\delta_{\mu,0}} J^{-1} a_\mu(-t,\bfx) J$, which indeed acts on the gauge indices.}
The $\T$ action commutes with the $PSp(N_f)$ global symmetry, since if $U\in PSp(N_f)$ then $(\unit_{N_f} \tp J)U = U^* (\unit_{N_f} \tp J)$, and therefore the $\unit_{N_f} \tp J$ and the complex conjugation involved in the $\T$ action cancel when acting on the $PSp(N_f)$ matrices. 

In addition to $\T$ we have parity, which is comparably simpler. We may take the action on the fermions to be 
\be \mcp: X(t,\bfx) \mt \gamma_0 X(t,-\bfx),\ee 
so that $\mcp^2 = (-1)^F$. The transformation of $a^I_\mu$ is dictated by that of $\p_\mu$, viz.
\be \ba \mcp : a^I_0(t,\bfx) & \mt a^I_0(t,-\bfx) \\ a^I_i(t,\bfx) &\mt -a^I_i(t,-\bfx).\ea\ee

Finally, we come to charge conjugation $\C$. Charge conjugation is often elevated to the same status as $\T$ and $\mcp$, but in fact there is no universal definition of a distinct $\C$ symmetry that applies in all theories.\footnote{At a formal level, it is usually only defined when one has a symmetry group that admits a non-trivial $\Z_2$ outer automorphism (an outer automorphism of a group $G$ is a homomorphism from $G$ to itself whose action cannot be written as conjugation by elements in $G$; complex conjugation for the group $U(1)$ is an example). For example, ${\rm Out}(SO(4N_f)) = \Z_2$, and so the un-gauged theory has a charge conjugation symmetry, viz. the reflection extending $SO(4N_f)$ to $O(4N_f)$. Likewise, ${\rm Out}(SU(N)) = \Z_2$ for $N>3$, with charge conjugation acting by exchanging the $SU(N)$ fundamental and anti-fundamental representations. However, ${\rm Out}(SU(2)) = {\rm Out}(PSp(N_f))$ is trivial, which precludes the existence of an independent notion of charge conjugation in the gauge theory under consideration.}
Indeed, in the $SU(2)$ gauge theory under consideration there is actually no notion of a charge conjugation symmetry which is distinct from the other symmetries already discussed. We will use the symbol $\C$ to denote the unitary map which acts on the complex fermions as $\C : \psi_{\sigma v} \mt \psi^*_{\sigma v}$; on the matrix field $X$ this action is $\C : X \mt (\unit_{N_f} \tp J)^{-1} X J$. This can be checked to be a symmetry of the action provided that $\C : a_\mu \mt J^{-1} a_\mu J$. The reason why this does not give us an independent symmetry is that it is already contained in the action of $PSp(N_f)\times SU(2)$, and so $\T$ and $\CT$ are related through the internal symmetry group. 
Therefore what we call $\T$ and what we might call $\CT$ are not distinct symmetries after the $\psp \times SU(2)$ has been properly accounted for, and we can restrict ourselves to just dealing with $\T$ without any loss of generality. 


Recapitulating, we have a $PSp(N_f)$ global flavor symmetry and discrete $\T$ and $\mcp$ symmetries, which act on the dynamical fields as
\be\ba\, PSp(N_f) :\,  & X \mt UX,\quad a^I_\mu\mt a^I_\mu,\\ \T  :\, & X(t,\bfx) \mt \gamma_0\gamma_5 X^*(-t,\bfx), \\ & a^I_0(t,\bfx) \mt -a^I_0(-t,\bfx),\\ & a^I_i(t,\bfx) \mt a^I_i(-t,\bfx),\\  \mcp  :\, & X(t,\bfx) \mt \gamma_0X(t,-\bfx), \\ & a^I_0(t,\bfx) \mt a^I_0(t,-\bfx),\\ & a^I_i(t,\bfx) \mt -a^I_i(t,-\bfx).\ea\ee
With these definitions, the Dirac mass $m\Tr[i\bar X X]$ is even under both $\T$ and $\P$, while the chiral mass $m\Tr[\bar X \gamma_5 X]$ is odd under both $\T$ and $\P$. 

\section{Discrete topological terms and fractional instanton numbers}
\label{app:instantons}

In this appendix we discuss discrete theta terms in four dimensions and their relation to instanton numbers, which is important for analyzing the possible presence of SPTs that survive $\T$ symmetry breaking in the theories discussed in the main text. 

A general framework for thinking about topological terms in gauge theories is that of obstruction theory. Given a gauge theory with gauge group $G$, we ask whether or not it is possible to have $G$-bundles over a given spacetime that cannot be trivialized, i.e. which do not admit a global section. If there is an obstruction to trivializing a given $G$-bundle $E$, then we can have non-trivial topological terms in the theory.
 
To determine whether a $G$-bundle can be trivialized over a given spacetime manifold $M$, it is helpful to imagine triangulating $M$. We first choose a 
trivialization (a choice of local section) at each of the vertices of the triangulation; this is always possible to do in a globally consistent way. On 
orientable manifolds, it is further always possible to extend this trivialization from the vertices onto the links. During the next step of extending the 
trivialization smoothly into the faces we may run into trouble, however. Indeed, suppose that $\pi_1(G) \neq 0$ and that around a given face, the trivialization on 
the links around the face determines a nontrivial element of $\pi_1(G)$. In this case, the trivializaton cannot be smoothly extended into this face, and we have an 
obstruction. 

At the next level up, we ask whether the trivialization can be extended into the 3-cells. All Lie groups have $\pi_2(G)=0$, and so such an extension is 
always possible. At the final level, we ask whether we can extend the trivialization into the 4-cells. All simple compact Lie groups have $\pi_3(G) = \Z$, 
and so we always run into an obstruction at this stage---this is the obstruction responsible for the usual instanton number.

Topological terms are functions which take information about the obstructions defined above and output a phase in the path integral, and they do so in such 
a way that the phase is invariant under re-triangulations of $M$. Terms in the topological part of the action come in two types: continuous $\theta$ terms 
which can appear with a continuously-variable coefficient in the action, and discrete terms, which are integrals of $\Z_n$-valued terms and have quantized 
coefficients. 

If there is an obstruction to extending the trivialization over the $k$-cells which comes from a factor of $\Z$ in $\pi_{k-1}(G)$, such as the instanton 
number when $k=4$, then a theta term associated with this obstruction may have a continuous coefficient, essentially because the representations of $U(1)$ are integers. However, if the obstruction comes from a finite group $\Z_m$, then the theta term must be discrete and come with a quantized coefficient, since it must assign a trivial phase to a certain number of copies of the gauge bundle 
$E$. Therefore discrete theta terms can only arise when there is torsion in the homotopy groups $\pi_{k<D}(G)$, with $D$ the spacetime dimension. Since 
$\pi_3(G) = \Z$ is always torsion-free and $\pi_2(G)$ always vanishes, the only place where discrete terms can possibly enter the game is at the level of 
the obstructions at the 2-cells.\footnote{We are grateful to helpful correspondence with Yuji Tachikawa on these issues.} If $\pi_1(G)$ contains a single 
$\Z_m$ factor (this will be true for all the groups we consider) then the resulting discrete $\theta$ term is constructed using a ``second Stiefel-Whitney 
class" $w_2$, which is a $\Z_m$-valued degree-2 characteristic 
class that measures the torsion part of the obstruction at the level of the faces of the triangulation \cite{scorpan2005wild}. The only way to get a four-dimensional topological term using just $w_2$ is to write 
\be S_{top} \supset  \frac{2\pi k}{2m}\int P(w_2),\ee
where $P(w_2)$ is the Pontryagin square, which for us is a 4-cochain valued in $\Z_{2m}$. $P(w_2)$ is the appropriate way of adapting the wedge product to discrete forms defined on a lattice, in that $P(w_2)$ measures the self-intersection number of the surface defined by $w_2$. The Pontryagin square of a sum factors in the way that we expect of a squaring operation\cite{kapustin2014topological}, namely $P(a+b) = P(a) + P(b) + 2 a\cup b$. Since $w_2$ is a $\Z_m$ class, we need $S_{top}$ to be invariant under the shift $w_2 \mapsto w_2 + m c$, where $c$ has integer periods. This requirement forces $km\in 2\Z$ \cite{zohar2017,cordova2019anomalies}. 
Therefore for the groups considered in this paper, the most general response we can consider is one parametrized by an angle $\theta$ and an integer $k$:
\be S_{top}[\theta,k] = \theta l + \frac{2\pi k}{2m} \int P(w_2),\ee
where the instanton number is $l = \int \frac{1}{8\pi^2} \Tr [F\wedge F] = \int \ch_2$, with $\ch_2$ the second Chern character of the bundle in question. 

We will always normalize the instanton number such that if $\widetilde G$ is the simply-connected universal cover of the gauge group $G$, then the minimal instanton for a $G$ bundle which lifts to a $\widetilde G$ bundle has $l=1$. This means that if $\pi_1(G) \neq 0$ so that $\widetilde G \neq G$, the instanton number in $G$ may be fractional, and consequently, the $\theta$ angle may have a periodicity greater than $2\pi$. 

For the groups we are interested in, the fractional part of the instanton number will be a function of $\int P(w_2)$. Depending on what this function is, it may be possible that the discrete $P(w_2)$ term in $S_{top}$ can be incorporated to the $\theta l$ term, a question which is addressed in detail in \cite{cordova2019anomalies, seiberg2013}. For example, suppose the fractional part of the instanton number can be written as $\frac{2\pi q}{2m}\int P(w_2)$, and for simplicity take $m\in 2\Z$ so that the distinct values of $q$ are given by $q\in\Z_{2m}$ ($\int P(w_2)$ can then generically take any value in $\Z_{2m}$ on a general non-spin manifold). If $\gcd(q,2m)=1$, then we see that any potential discrete theta term in $S_{top}[\theta,k]$ can be absorbed into the instanton term, via \be S_{top}[\theta,k] = S_{top}[\theta + 2\pi r,0],\ee where $r = k q^{-1}$, with the inverse taken in $\Z_{2m}$. 
If this happens then we may write $S_{top}$ entirely in terms of a term with a coefficient $\theta$ that may be continuously tuned, and hence upon breaking $\T$, the topological response is not protected by any symmetry, and may be continuously tuned away. On the other hand if $\gcd(q,m)>1$, then there are some discrete terms which cannot be re-expressed as a continuous theta term, and a protected topological response remains even after breaking $\T$.

To examine when this can happen, we must then compute the fractional part of the instanton number as a function of $P(w_2)$. In what follows we will be rather didactic and show in detail how this can be done for the case of $PSp(n)$ by following the approach in following the method of Refs. \onlinecite{witten2000supersymmetric} and  \onlinecite{cordova2019anomalies}, which interested readers can see for further examples. 

In order to get a fractional instanton number, we need to consider a $PSp(n)$ bundle $E$ which does not lift to an $Sp(n)$ bundle. This will be the case if the transition functions between patches in $E$ fail the cocycle condition by $-\unit_{2n} \in Z(Sp(n))$ along a collection of triple patch overlaps that defines a homologically non-trivial 2-manifold, whose Poincare dual is the $\Z_2$-valued class $w_2$. 

To construct such a bundle, consider the bundle $E_{SO(3)} = L^{1/2} \oplus L^{-1/2}$, which is an $SO(3)$ bundle that does not lift to an $SU(2)$ bundle. Here $L$ is a line bundle whose first Chern class reduces mod $2$ to the class $w_2$, so that $L^{1/2}$ is a line bundle with fractional flux, whose transitions fail the cocycle condition by $-1$'s in a way determined by $w_2$ (the opposite powers $\pm 1/2$ appearing in $E_{SO(3)}$ are needed so that $E_{SO(3)}$ has zero first Chern character, as required of any $SO(3)$ bundle). 

In order to make a $PSp(n)$ bundle, we then use the diagonal embedding $SU(2) \to Sp(n)$ to form the bundle 
\be E_{PSp(n)} = E_{SO(3)}^{\oplus n} =(L^{1/2} \oplus L^{-1/2})^{\oplus n}.\ee
Because of the direct sum, the transition functions in $E_{PSp(n)}$ fail the cocycle condition by $-\unit_{2n}$ in a way controlled by the class $w_2$, which is what we want.\footnote{If we had not used the diagonal embedding of $SU(2)$ we would have produced a bundle whose transition functions failed the cocycle condition in a way not proportional to the identity matrix, which is not allowed. }

The instanton number $l$ of $E$ is the integral of the second Chern character $\ch_2(E)$. 
This can be computed using the relation $\ch_2(A \oplus B) = \ch_2(A) + \ch_2(B)$, so that 
\be \ch_2(E_{PSp(n)}) = n (\ch_2(L^{1/2}) + \ch_2(L^{-1/2})).\ee
Now since $L^{\pm 1/2}$ is a line bundle, $\ch_2(L^{\pm 1/2}) = \frac{1}{2} \ch_1(L^{\pm 1/2})\wedge \ch_1(L^{\pm 1/2})$. The first Chern character of $L^{\pm 1/2}$ has an integer part and a fractional part, with (by definition) the fractional part given by $\pm w_2/2$. Accounting for the fact that the correct way of taking the wedge product when discrete terms are involved is to use the Pontryagin square so as to properly capture the intersections of the surfaces Poincare dual to $w_2$, we find that in general 
\be l = \int  \ch_2(E_{PSp(n)}) = \frac{n}{4} \int P(w_2) + \dots,\ee
where the $\dots$ represent integer-valued terms coming from the "small" instantons which are also present for $Sp(n)$ bundles. The consequence of this is that when $n\in 2\Z+1$, we may have fractional instantons on any manifold. Since $P(w_2)/2$ is always an integer class on a spin manifold, if $n \in 2\Z$ we may have fractional instantons on non-spin manifolds only, while if $n \in 4\Z$, fractional instantons never appear.
 From this expression we also see that when $n$ is odd, the discrete term $2\pi k/4 \int P(w_2)$ in $S_{top}[\theta,k]$ can always be absorbed into a shift of the continuous $\theta$ term, while for $n$ even it cannot be. 

Now we briefly address the possibility of having fractional instantons in the $SU(N_c>2)$ gauge theory case, where the global flavor symmetry is $U(N_f)/\Z_{N_c}$. As discussed above, in order to determine what discrete theta terms are are possible, we need to determine the fundamental group $\pi_1[U(N_f)/\Z_{N_c}]$. One can do this by examining the long exact sequence in homotopy groups stemming from the 
short exact sequence $1 \to \Z_{N_c} \to U(N_f) \to U(N_f)/ \Z_{N_c}\to 1$. Since $\pi_1[U(N_f)] = \Z$ for all $N_f$, the relevant part of the homotopy group sequence is 
\be 1 \to \Z \xrightarrow{\pi} \pi_1[U(N_f)/\Z_{N_c}] \to \Z_{N_c} \to 1.\ee 
The map $\pi$ can be determined by examining how the minimal non-contractible loop in $U(N_f)$ maps to $U(N_f)/\Z_{N_c}$. Doing this fixes the central term of the sequence to be 
\be \pi_1[U(N_f)/\Z_{N_c}] = \Z \times \Z_{\gcd(N_f,N_c)},\ee
which means in particular that $\pi_1[U(N_f)/\Z_{N_c}]=\Z$ is torsion-free when $N_c$ and $N_f$ are relatively prime. Intuitively, this is because in this case, the quotient cannot hit the $SU(N_f)$ factor in $U(N_f)$, which is the only place torsion can come from, on account of $\pi_1[SU(n) / \Z_m] = \Z_m$ for $m$ dividing $n$. 

Hence when $\gcd(N_f,N_c)=1$ (which includes the
situations we are interested in, viz. odd $N_f$ and (perhaps large) $N_c$), no discrete topological response is possible. More generally, one can use the methods described in this appendix (namely constructing bundles with fractional instanton numbers out of direct sums of fractional line bundles) to show that no indpendent discrete response is possible provided that
\be  \label{ungcd} \gcd \( \frac{N_f(N_f-1)}{g}, 2g\) = \begin{cases} 1 & g \in2\Z \\ 2 & g \in 2\Z+1\end{cases},\ee
with $g\equiv \gcd(N_f,N_c)$. 

\section{A lattice model for the parton theory} 
\label{app:lattice}
Here we present an explicit UV regularization for the $SU(2)$ gauge theory with massless Dirac fermions on the lattice. We start with a lattice model of a single low energy Dirac fermion. Let us consider a $3d$ cubic lattice with 4 orbitals on each site. The Hamiltonian in momentum space is given as the following:
\bea
H&=&\sum_{\bm{k}}c_{\bm{k}}^\dagger\left(\sum_{i={x,y,z}}\mu^z\otimes\sigma^i \sin k_i\right) c_{\bm{k}}\\
&+& \sum_{\bm{k}}c_{\bm{k}}^\dagger\left(m_0\mu^x+m_1\mu^x\sum_{i={x,y,z}}\cos k_i\right) c_{\bm{k}},
\eea
where $\mu$'s and $\sigma$'s are pauli matrices acting on the orbital basis. For the lattice model, we can define the $\mathcal{C}$, $\T$ and $\P$ symmetries which act as 
\bea
\mathcal{C}&:& c_{\bm{k}}\rightarrow \mu^{y}\otimes \sigma^yc_{-\bm{k}}^\dagger,\\
\mathcal{T}&:& c_{\bm{k}}\rightarrow i\sigma^{y}c_{-\bm{k}},\ \ \ i\rightarrow -i,\\
\mathcal{P}&:& c_{\bm{k}}\rightarrow \mu^{x}c_{-\bm{k}}.
\eea

At $m_0=-3m_1$, this model has a single Dirac fermion at $\bm{k}=(0,0,0)$. The effective Hamiltonian for this fermion is simply
\beq
H_{Dirac}=\int_{\bm{k}}\psi_{\bm{k}}^\dagger\left(\sum_{i={x,y,z}}\mu^z\otimes\sigma^i  k_i\right) \psi_{\bm{k}}.
\eeq
The lattice $\mathcal{C}$,
$\mathcal{T}$ and $\mathcal{P}$ symmetries precisely map to the continuum $\mathcal{C}$,
$\mathcal{T}$ and $\mathcal{P}$ symmetries for the low energy Dirac fermion without additional complication. The fermion mass term $\psi^\dagger\mu^x\psi$ is the only allowed fermion bilinear term that preserves these discrete symmetries. 

We can add additional flavor indices to the above model. In particular, we may take $2N_f$ flavors of this model and then couple it to lattice $SU(2)$ gauge fields in the standard fashion. This serves as a UV regularization for the parton field theory in Eq. \ref{eq:L}. The combination $\mathcal{CT}$ is mapped to the $\T$ symmetry we defined in the parton theory Eq. \ref{eq:L}. The $\mathcal{C}$ symmetry becomes part of the continuous $PSp(N_f)$ symmetry in the parton theory.

\section{Counterexample: $1d$ Schwinger model} 
\label{app:1dSchw}
In this appendix we discuss the Schwinger model, which is 1+1-D QED with $N_f\in 2\Z+1$ Dirac fermions: 
\beq
\mathcal{L}_{1d}=i\sum_{i=1}^{N_f}\left(\bar{\psi}_i\gamma_\mu(\partial_\mu-ia_\mu)\psi_i-m\bar{\psi}_i\psi_i\right)+\frac{1}{4e^2}f^2+...
\label{eq:1d}
\eeq
At first sight, the physics of this model parallels that of the $3+1$-D theories encountered in the main text: this $1+1$-D theory is also intrinsically bosonic, and the system will go through a spontaneous $\mathcal{P}$ breaking transition as the mass is tuned from negative to positive. However, we will show that the symmetry breaking states in the end are topologically distinct from the trivial state, which is different from the $3+1$-D examples in the main text. 

Let us consider $N_f=3$ for simplicity. The important part of the global symmetry of the theory is $PSU(3)\times \Z_2^\P$, where $\mathcal{P}$ is the discrete parity symmetry.
These symmetries are enough to prevent other fermion bilinear terms.\footnote{In the real basis $\gamma_0=-i\sigma^y, \gamma_1 = \sigma^x$, we take $\P:\psi \mt \gamma_0\psi$ and $\T:\psi \mt \bar\psi \sigma^z$, with $\P^2 = (-1)^F$ and $\T^2 = \unit$. Both $\P$ and $\T$ preserve the Dirac mass $im\bar\psi_i\psi_i$, while the chiral mass $\bar\psi_i \gamma_5\psi_i$ is odd under both $\P$ and $\T$. We also have the unitary operator $\C$, which in our basis just does $\C : \psi \mt\psi^*$.} 
At the $m=0$ point, the IR physics of the model is described by the $SU(3)_1$ WZW conformal field theory. By turning on the fermion mass, we can get two massive phases. On the $m<0$ side, we choose the regularization such that the $U(1)$ gauge theory has $\theta=0$. 
This phase has a gapped non-degenerate ground state that does not break any symmetry. On the $m>0$ side the $\theta$-angle is $\theta=3\pi$. The states of $1+1$-D $U(1)$ gauge theory with a theta term have energies determined by the Hamiltonian $H_{eff}\sim E^2 = (n-\frac{\theta}{2\pi})^2$ 
with $n\in \Z$. For $\theta=3\pi$, the theory has two-fold degenerate ground states, namely $E=\pm\frac{1}{2}$, which spontaneously break $\mathcal{P}$ on account of $\P : E \mapsto -E$. 
The relative topology we claimed above between two states and the trivial state can be understood in the following way. We can start from a system with $\theta=0$ and adiabatically tune up $\theta$ to $3\pi$. 
Correspondingly, the electric field strength $E$ will adiabatically increase to $3/2$. 
To get to the $E=+1/2$ state, a pair of $\pm 1$ charges must be nucleated and sent to the boundaries of the system, while a pair of $\pm 2$ charges is required to get to the $E=-1/2$ state. 
Since the only charge $\pm1,\pm 2$ objects of the system carry projective representation of the $PSU(3)$ symmetry, the $E=\pm 1/2$ states differ from trivial state by an SPT state protected by $PSU(3)$. The $E=1/2$ and $E=-1/2$ states themselves also differ by a $PSU(3)$ SPT, for the same reason.

 \begin{figure}
\begin{center}
\includegraphics[width=0.9\linewidth]{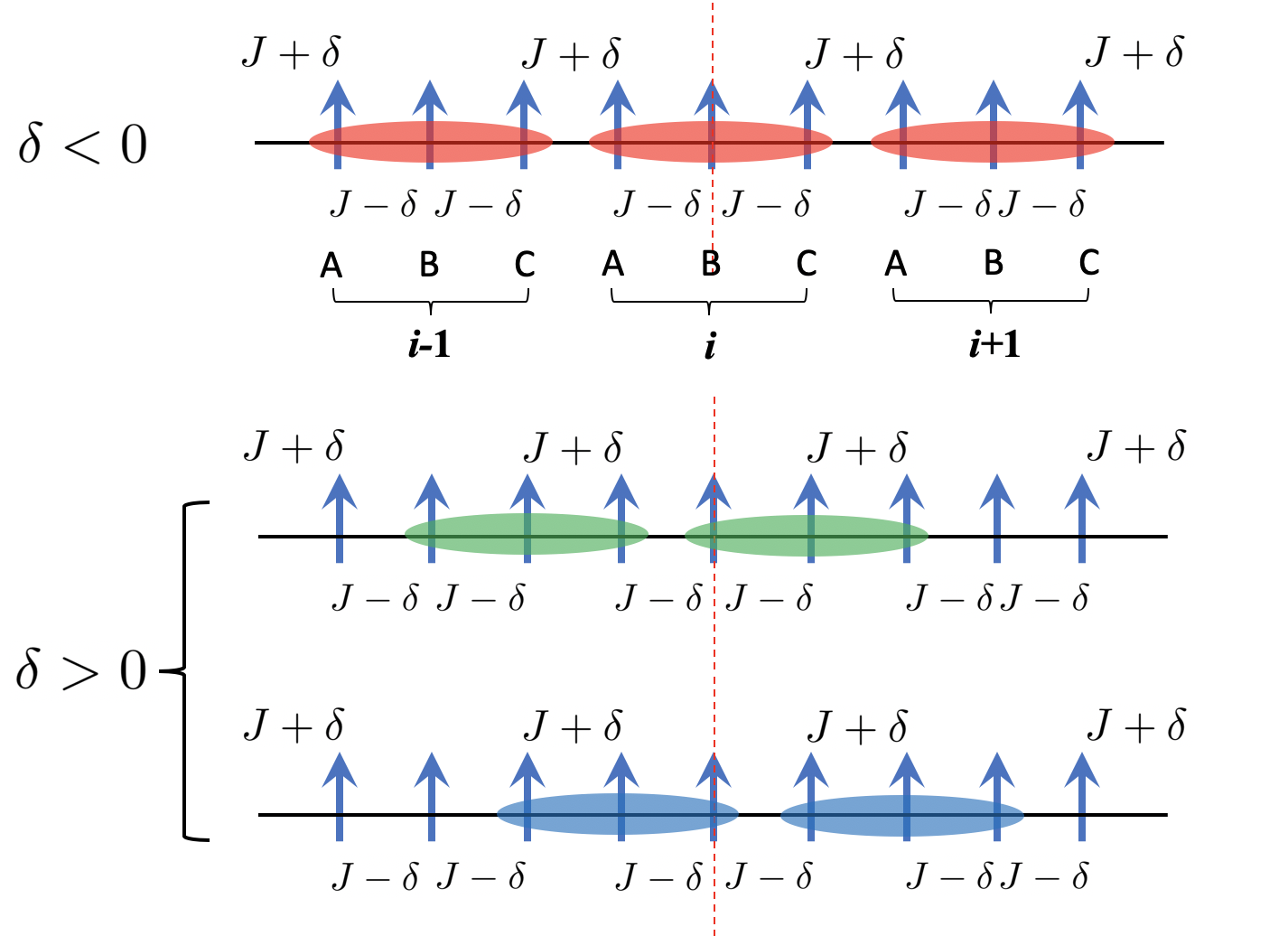}
\includegraphics[width=0.9\linewidth]{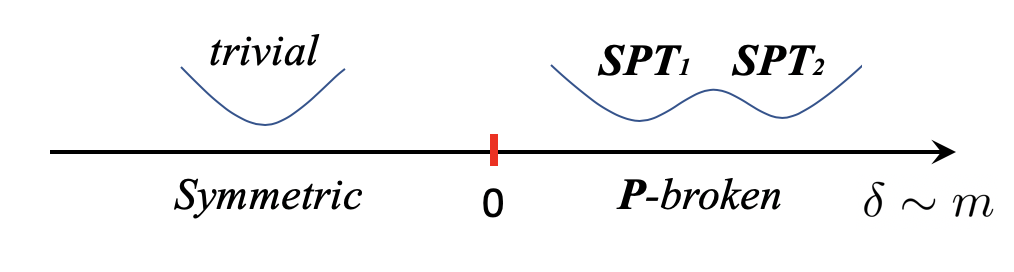}
\caption{1d $SU(3)$ spin chain with fundamental spin at each site. The nearest neighbor spins interact with each other through a Heisenberg Hamiltonian with interaction strength labeled in the figure. For $\delta\neq 0$, one unit cell contains three spins. For $\delta=0$ the system has enlarged translational symmetry. The Hamiltonian has a site-centered parity symmetry. }
\label{fig:1d}
\end{center}
\end{figure}

The above argument can be made precise by an explicit lattice model. Consider an $SU(3)$ spin chain with fundamental representation on each site. The Hamiltonian is written as (also see Fig. \ref{fig:1d})
\bea
\nonumber
H_{spin}&=&(J-\delta)\sum_{i}\left( S_{i,A}\cdot S_{i,B} +S_{i,B}\cdot S_{i,C} \right)\\
&+& (J+\delta)\sum_{\langle i,j\rangle} S_{i,C}\cdot S_{j,A},
\eea
where $S_i\cdot S_j$ is a short hand notation for $S_{\alpha}^\beta(i)S_{\beta}^{\alpha}(j)$. The spin model at $\delta=0$ realizes the $SU(3)_1$ conformal field theory, which is exactly the theory in Eq. \ref{eq:1d} at $m=0$. Eq. \ref{eq:1d} can be viewed as a parton mean field theory of the spin chain\cite{spinchain}. One can show that the $\delta$ term in the spin chain precisely maps to the fermion mass term in the parton theory, namely $\delta\sim m$. For $\delta<0$, the spin chain has a unique ground state which is a tensor product of trimers formed between spins on ABC sublattices within a unit cell. This corresponds to the $m<0$ phase in the parton theory. For $\delta>0$, there are two degenerate trimerization patterns as shown in Fig. \ref{fig:1d}, which corresponds to the two-fold ground states in the $m>0$ phase of Eq. \ref{eq:1d}. The two patterns, as shown in Fig. \ref{fig:1d}, leave some boundary spins unpaired, similar to the boundary state of AKLT chain. Therefore, they are topologically distinct from the $\delta<0$ state, and hence the critical theory does not describe a conventional Landau ordering transition.

\section{Continuous Landau-forbidden transitions in $3+1$-dimensions}
\label{app:3dsu2b}
Here we display a bosonic model that shows a continuous phase transition with a deconfined critical point between two Landau-allowed phases that break distinct symmetries. This is thus a direct analog in $3+1$-D of the phenomena discussed previously\cite{2dDQCP1,2dDQCP2,2dDQCPjps,2dDQCPprb,kaul2013bridging,sandvik2010} in $2+1$-D. 
Consider $SU(2)$ gauge theory at $\theta = \pi$ coupled to $N_b$ bosons in the fundamental representation of the $SU(2)$ gauge group: 
\be \ba \mathcal{L} & = \frac{1}{4g^2}\Tr\,f_{\mu\nu}^2 + \frac{\pi}{8\pi^2} \Tr(f\wedge f) \\
&\quad  + |(\p_\mu - ia^I_\mu\sigma^I/2)\phi|^2 - r |\phi|^2 - \frac{\lambda}{2}|\phi|^4.\ea\ee
(The $I$ index runs from 1 to 3). 
Here time reversal symmetry acts as 
\be \ba \T   :\, &  \phi(t,\bfx) \mt \phi^*(-t,\bfx)\\ 
&  a^I_0(t,\bfx) \mt -a^I_0(-t,\bfx), \\ &  a^I_j(t,\bfx) \mt a^I_j(-t,\bfx).\ea\ee

The $\theta$ term can be viewed as arising from  a distinct set of heavy `spectator' bosons $\phi_H$ (also in the fundamental of $SU(2)$) that form a boson SPT phase of the $SU(2)$ group before it's gauged. We will take $\phi_H$ to transform in the same way as $\phi$ under time reversal but to be a flavor singlet. 
The global symmetries of this theory are almost identical to the theory with fermionic matter discussed in the main text but with one difference. The presence of the flavor singlet  spectator boson $\phi_H$ implies that there are gauge-invariant operators $\phi_H^\dagger \phi$ that transform in the fundamental representation of $Sp(N_b)$. Thus the continuous global symmetry is $Sp(N_b)$ and not $PSp(N_b)$. From a formal point of view, in the presence of the spectator bosons, the $SU(2)$ gauge bundle and the background $Sp(N_b)$ bundle are independent of each other (the condition in Eq. \ref{sw_constraint} does not hold). We also assume that the spectator bosons do not contribute a nontrivial SPT of the global symmetry.

Since a Dirac fermion behaves like four bosons for the purposes of computing the flow of the gauge coupling, the 1-loop beta functions\footnote{The pertubative RG is presumably not affected by the theta term of the $SU(2)$ gauge field, whose non-perturbative effects are are exponentially suppressed at weak coupling.} for $\bar g^2 \equiv g^2/(8\pi^2)$ and $\bar\lambda\equiv \lambda /(8\pi^2)$ at the massless point are
\be \ba   \frac{d\bar g^2}{dl} & = \left( \frac{22}{3} - \frac{1}{6}N_b\right) \bar g^4, \\
\frac{d\bar\lambda}{dl} & = -(2N_b+4)\bar\lambda^2 + \frac{9}{2}\bar g^2 \bar\lambda - \frac{9}{8}\bar g^4,
\ea\ee
which can be extracted from Ref. \onlinecite{gross1973asymptotically}. 
For $N_b > 359$, 
and when the boson mass is tuned to zero, the gauge theory becomes IR free, while for $N_b < 359$ the theory has an instability towards $\bar\lambda<0$, signaling a first-order transition.  
The zero boson mass point, for $N_b>359$ (which we will assume in what follows), separates a Higgs phase where the bosons condense to one in which the bosons are gapped. 

\begin{figure}
\begin{center}
\includegraphics[width=0.9\linewidth]{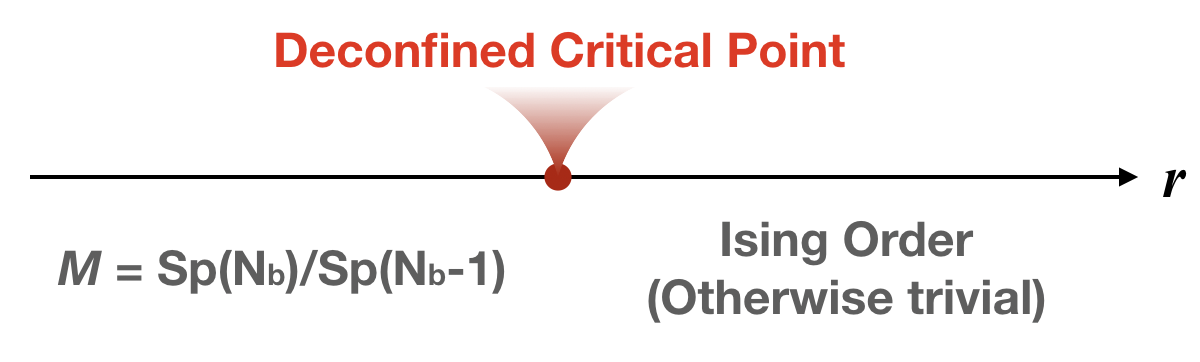}
\caption{A schematic phase diagram for the continuous Landau-forbidden transition in $3+1d$. }
\label{fig:3d}
\end{center}
\end{figure}

When the bosons condense, the $SU(2)$ gauge field is fully Higgsed and there is no residual unbroken gauge structure.  The $Sp(N_b)$ symmetry is broken to $Sp(N_b -1)$ while preserving time reversal. This is a conventional Landau ordered phase characterized by a gauge invariant order parameter $Tr(\phi \phi^\dagger)$. On the side of the phase transition,  the bosons are gapped. Then the low energy physics is described by $SU(2)$ gauge theory at $\theta = \pi$ which possibly breaks time reversal but preserves $Sp(N_f)$.  The discussion  in the main text about the lack of any SPT order protected by the unbroken symmetry generalizes to the present situation as well. Thus this phase is also a conventional Landau ordered phase with an Ising order parameter that captures the time reversal breaking.

As in the main text the theory can be generalized to $SU(N_c)$ or $Sp(N_c)$ where the $\mathcal{T}$ breaking is known with more confidence. The two phases then break distinct symmetries  but are both  Landau allowed. A continuous phase transition between these two phases is forbidden within Landauesque thinking but is possible through the deconfined critical route just described. 

\newpage

\bibliography{LBL.bib}

\end{document}